\documentclass[11pt,a4paper]{article}
\usepackage{jcappub}

\usepackage{bm}
\usepackage{epsfig}
\usepackage{citesort}
\usepackage{graphicx}
\usepackage{amssymb}
\usepackage{color}
\usepackage{subfigure}
\usepackage{slashed}
\usepackage{afterpage}
\usepackage{psfrag}

\renewcommand\({\left(}
\renewcommand\){\right)}
\renewcommand\[{\left[}
\renewcommand\]{\right]}
\newcommand{\be}{\begin{equation}}
\newcommand{\ee}{\end{equation}}
\newcommand{\bea}{\begin{eqnarray}}
\newcommand{\eea}{\end{eqnarray}}
\newcommand{\parte}{{\ttfamily PArthENoPE}}

\newcommand{\exclude}[1]{}

\renewcommand\({\left(}
\renewcommand\){\right)}
\renewcommand\[{\left[}
\renewcommand\]{\right]}

\newcommand{\decay}{\Gamma_{\gamma}}

\begin{document}
\subheader{\hfill MPP-2011-116}

\title{Cosmological bounds on\\
pseudo Nambu-Goldstone bosons}

\author[a]{Davide~Cadamuro}
\author[a]{Javier~Redondo}

\affiliation[a]{Max-Planck-Institut f\"ur Physik
(Werner-Heisenberg-Institut)\\
F\"ohringer Ring 6, D-80805 M\"unchen, Germany}

\emailAdd{cadamuro@mppmu.mpg.de}
\emailAdd{redondo@mppmu.mpg.de}

\abstract{We review the cosmological implications of a relic population of pseudo Nambu-Goldstone bosons (pNGB) with an anomalous coupling to two photons, often called axion-like particles (ALPs). 
We establish constraints on the pNGB mass and two-photon coupling by considering big bang nucleosynthesis, the physics of the cosmic microwave background, and the diffuse photon background.
The bounds from WMAP7 and other large-scale-structure data on the effective number of neutrino species can be stronger than the traditional bounds from the primordial helium abundance. 
These bounds, together with those from primordial deuterium abundance, constitute the most stringent probes of early decays.}

\maketitle

\section{Overview}\label{sec:motivation}

The spontaneous symmetry breaking (SSB) of exact global continuous symmetries produces \emph{massless} Nambu-Goldstone bosons. 
If the symmetries are only approximate, the putatively massless bosons acquire small masses and are called \emph{pseudo} Nambu-Goldstone bosons (pNGB). 
The axion is a well studied example of the latter: it arises from the SSB of the Peccei-Quinn U(1)$_{\rm PQ}$ axial symmetry, 
postulated to solve the strong CP problem \cite{Peccei:2006as,Nakamura:2010zzi}. 
This symmetry is explicitly broken only by the colour anomaly. The axion potential becomes $\sim \Lambda_{\rm QCD}^4\cos(a/f_a)$, 
where $a$ is the axion field,  $f_a$ is the PQ SSB scale up to an integer and $\Lambda_{\rm QCD}$ is the confinement scale of quantum chromodynamics.
As a consequence the axion picks up a mass\footnote{In nature there is another pNGB, the $\eta'$ meson (related to a common axial phase shift transformation of the light quarks) with which the axion mixes and through it, with the rest of pseudoscalar mesons. 
At the end of the day, the mass of the axion further reduces to $m_a\sim m_\pi f_\pi/f_a$.} 
$m_a\sim \Lambda_{\rm QCD}^2/f_a$.
At energies much below $f_a$, the axion field always enters into the effective
Lagrangian in the combination $a/f_a$ so its interactions with standard model (SM) particles are
always suppressed by the high energy scale $f_a$. 

The axion can be generalised to a generic pNGB, which in this paper
we call an axion-like particle (ALP) and denote by $\phi$.
If the dynamics explicitly breaking the associated global continuous symmetry have a
characteristic scale $\Lambda$, the ALP mass is 
$m_\phi\sim \Lambda^2/f_\phi$, where $f_\phi$ is the scale at which 
the ALP symmetry is spontaneously broken. 
Interactions of ALPs with SM particles are also suppressed by this scale $f_\phi$. 
The phenomenology of the SM requires $\Lambda$ to be related
to physics beyond the electroweak scale, i.e.~$\Lambda\gtrsim$ TeV (which implies $m_\phi \gg m_a$ for  $f_\phi=f_a$), or to belong to a hidden sector. 
Axions, ALPs and other low mass particles postulated as a bridge between the hidden sector and the SM are typical examples of weakly interacting slim particles (WISPs), which are subjects of growing interest in the  low-energy high-precision frontier of particle physics~\cite{Jaeckel:2010ni}.

Axions and ALPs are subject to very strong constraints from stellar evolution and cosmology. 
The more stringent astrophysical bounds rely on the production of these exotic particles in the
hot and dense stellar interiors and their subsequent escape contributing directly to stellar energy losses
and therefore to the consumption of nuclear fuel. 
Normally the weaker the ALP interacts, the smaller the ALP production and 
the impact on stars.

In cosmology, the situation is reversed. Assuming sufficiently high initial
temperature of the universe, a thermal population of ALPs is created and decouples from the SM thermal bath at some point.  
The impact on the cosmological observables at our disposal---big bang
nucleosynthesis (BBN), anisotropies in the cosmic microwave background (CMB), 
etc.---mainly depends on when this relic population decays (if it decays at all). 
The decay lifetime is proportional to $f_\phi^2/m_\phi^3$, so the weaker the coupling, the slower the decay,  and the greater the relevance to late cosmology. 
Considering this, cosmological arguments are complementary to stellar evolution constraints and laboratory experiments.

In a recent paper~\cite{Cadamuro:2010cz}, we studied the constraints that
current cosmology places on cosmologically unstable axions (those with masses in the 
range of a few eV to MeV). In this paper we extend this study to a general ALP.
In~\cite{Cadamuro:2010cz} we found that the phenomenology is basically
determined by the anomaly-driven two-photon coupling, which for an ALP we write as 
\be\label{2phot}
\mathcal{L}_{\phi\gamma \gamma}=-\frac{g}{4} \,
F_{\mu\nu}\widetilde{F}^{\mu\nu}\, \phi
\ee
where $F_{\mu\nu}$ and $\widetilde F^{\mu\nu}$ are the electromagnetic field tensor
and its dual. 
The coupling strength $g\sim \alpha/(2\pi f_\phi)$ has inverse mass dimension and provides the ALP with an efficient thermalization mechanism in the early universe, i.e.~the Primakoff process $\gamma+q\to \phi+q$ ($q$ stands for any charged particle), and a prominent decay channel, i.e.~$\phi\to\gamma\gamma$. 

For most of the paper, we focus on this coupling, which we take as the defining characteristic of an ALP, together with its mass. 
The effects produced by other couplings are finally discussed in Sec.~\ref{sec:beyond}.

Alongside the thermal ALP population, a non-thermal population of ALPs can be created by the realignment mechanism just as in the axion case~\cite{Sikivie:2006ni}. The consequences of this population will be presented in a separate publication~\cite{Arias:2012mb}.

A number of excellent papers have considered the cosmological implications of decaying  massive relics,  
see for instance~\cite{Adhya:2003tr,Ellis:1990nb,Cyburt:2002uv}. 
Our work focuses on ALPs and deals with particles in an intermediate mass range (keV--GeV),
which is often not considered. 
As an additional complication, in a part of parameter space the inverse decay $\gamma\gamma\rightarrow\phi$ is also effective. ALPs might not completely decouple nor be in thermal equilibrium so we must follow the evolution of the ALP phase space distribution. 
This paper updates and complements the previous work of Mass\'o and Toldr\`a on a spinless particle coupled to photons~\cite{Masso:1995tw,Masso:1997ru}.

\begin{figure}[tbp]
\begin{center}
\psfrag{g}{$g$}
\includegraphics[width=13cm]{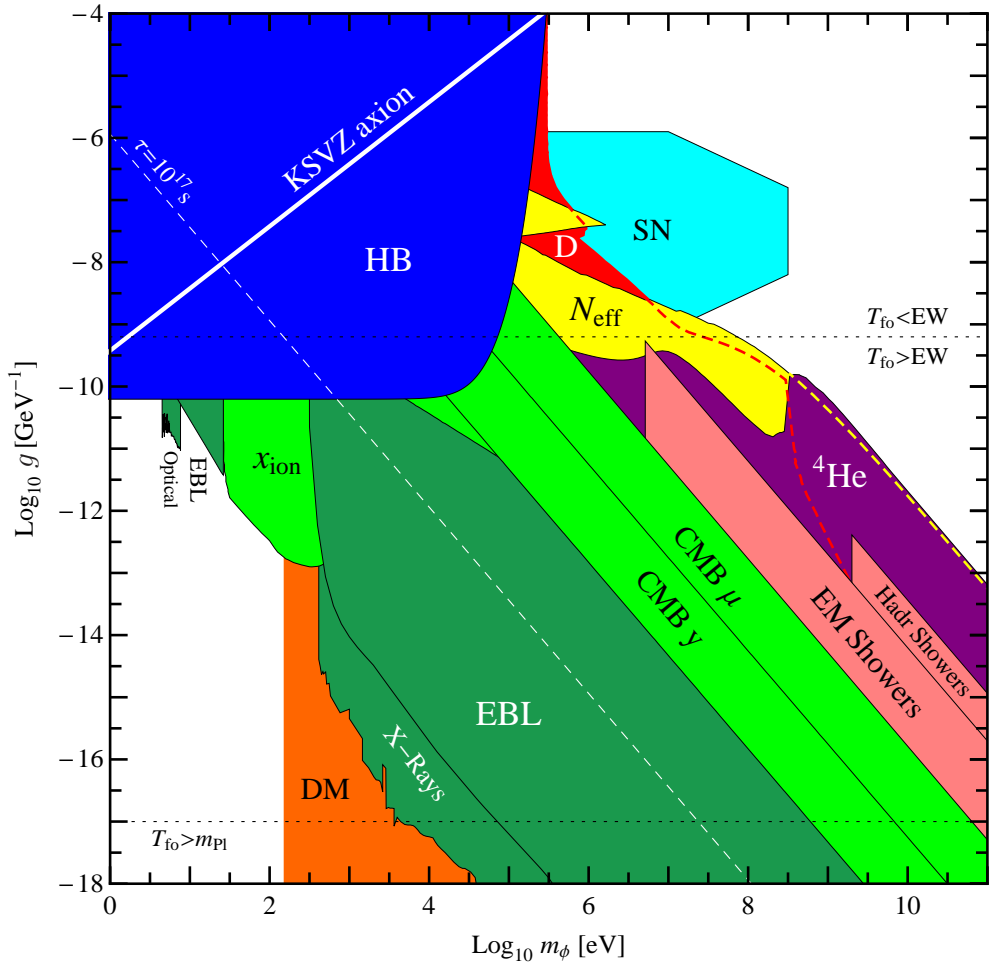}
\end{center}
\vspace{-.6cm}
   \caption{ALP bounds in the $m_\phi-g$ parameter space. 
   The labeling is described in Sec.~\ref{sec:motivation}.}
   \label{fig:bounds}
\end{figure}

\begin{figure}[tbp]
\begin{center}
\includegraphics[width=13cm]{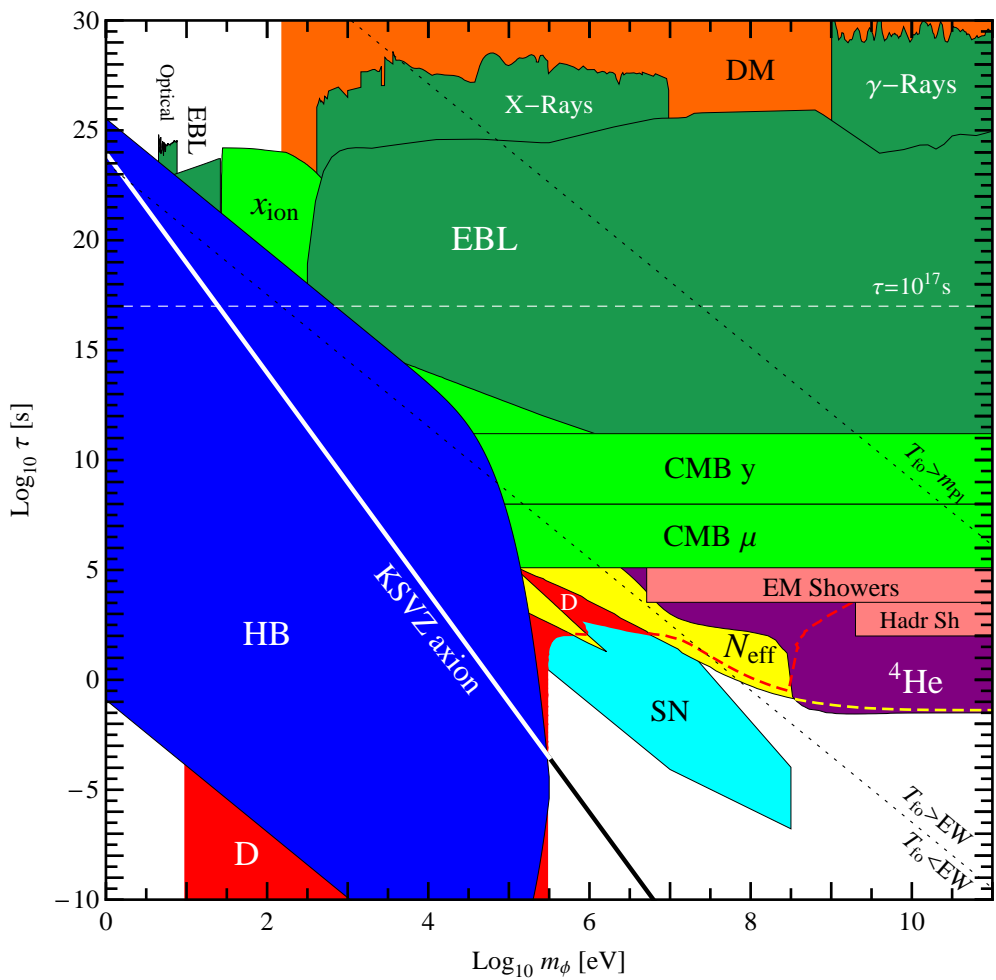}
\end{center}
\vspace{-.6cm}
   \caption{ALP bounds in the $m_\phi-\tau$ parameter space. 
   The labeling is described in Sec.~\ref{sec:motivation}.}
   \label{fig:boundsLife}
\end{figure}

Our conclusions are presented in Figs.~\ref{fig:bounds} and~\ref{fig:boundsLife}, which shows the constraints on ALPs 
coupled to photons. Cosmological bounds in Fig.~\ref{fig:bounds} have a characteristic slope $g\propto m_\phi^{-3/2}$ because they critically depend on the ALP lifetime 
\be
\tau \equiv \decay^{-1}= \frac{64 \pi }{m_\phi^3 g^2} 
\ee
and have a milder dependence on other parameters. 
Ordered by decreasing lifetimes, the excluded regions are: 
\begin{itemize}
\item {\bfseries DM} --- if ALPs are cosmologically stable and behave as dark matter (DM) they should not exceed the DM fraction measured by WMAP;
\item {\bfseries Optical, X-Rays, $\gamma$-Rays} --- photons produced in ALP decays inside galaxies would show up as a peak in galactic spectra that must not exceed the known backgrounds;
\item {\bfseries EBL} --- photons produced in ALP decays when the universe is  transparent must not exceed the extragalactic background light (EBL); 
\item {\bfseries $\mathbf{x_{\rm ion}}$} --- the ionization of primordial hydrogen caused by the decay photons must not contribute significantly to the optical depth after recombination;
\item {\bfseries CMB y, $\boldsymbol\mu$} --- if the decay happens when the universe is opaque, the decay photons must not cause spectral distortions in the CMB spectrum that cannot be fully rethermalized;
\item {\bfseries EM, Hadr showers} --- the decay of high mass ALPs produces electromagnetic and hadronic showers that must not spoil the agreement of big bang nucleosynthesis with observations of primordial nuclei; 
\item {\bfseries $\mathbf{^4}$He, D} --- the ALP decays produce photons (entropy) that dilute the baryon and neutrino densities, whose values affect the outcome of BBN, in particular the deuterium and $^4$He yields. Again, this dilution should not compromise BBN;
\item {\bfseries $\mathbf{N_{\rm eff}}$} --- the neutrino density must not disagree with the value measured by WMAP and other large-scale-structure probes. 
Currently, data points to a number of effective neutrinos $N_{\rm eff}$ greater than 3, 
which is disfavoured in the decaying ALP cosmology.
\end{itemize}

These bounds are complemented at low masses and large couplings by stellar-evolution arguments and laboratory searches. The most relevant astrophysical bounds come from star counts in the
colour-magnitude diagrams of globular clusters, in particular through the effect on the evolution of horizontal branch stars.
This constraint overlaps significantly with the cosmological  bounds, so we have computed precisely its turn off at large ALP masses (region labelled {\bfseries HB} in Figs.~\ref{fig:bounds} and~\ref{fig:boundsLife}).
The duration of the neutrino pulse from supernova 1987a can also constrain ALPs ({\bfseries SN})~\cite{Masso:1995tw}. 
However this bound relies on very sparse data and insufficient understanding of supernova dynamics and  ALP emission from a nuclear-density environment, which makes it somehow unreliable. 
This is a pity because the temperature and density of the SN core at the time of the collapse are among the highest we can find in stellar environments, so the Boltzmann suppression of the bounds appears only at larger masses~\cite{Raffelt:2006cw}. 

Much stronger couplings are tested by direct searches in laboratories (for a review see~\cite{Redondo:2010dp,Jaeckel:2010ni}), however the sky is the fundamental tool to test weakly coupled ALPs.

We have divided this paper in three main blocks. In Sec.~\ref{sec:relicabu} we discuss the relic abundance of ALPs. In Sec.~\ref{sec:early} we consider the cosmological implications of ALP decays before recombination, in the opaque universe. 
The bound from the effective number of neutrinos is the most relevant, since it covers most of the parameter space and constitutes the most relevant original 
contribution of this paper to the ALP parameter space. We also describe the 
implications on BBN. 

We describe the signatures and bounds on ALPs decaying after recombination in Sec.~\ref{sec:late}. We present bounds from the non observation of the decay photons and the ionization history of the universe and the possible signatures of ALP decay in the CMB spectrum. 
Finally, in Sec.~\ref{sec:beyond} we briefly analyse how our bounds change with the addition other couplings besides the two-photon one.
The computation of the high mass boundary of the HB constraint is discussed in the Appendix. 
The summary plots are Figs.~\ref{fig:bounds} and~\ref{fig:boundsLife} and our conclusions are drawn in Sec.~\ref{sec:end}.

\section{Establishment and decay of a thermal ALP population} \label{sec:relicabu}

The two-photon coupling, Eq.~\ref{2phot}, allows the establishment of a thermal
population of ALPs in the early universe through the Primakoff process. 
The rate of ALP production due to scattering on relativistic electrons was
computed in~\cite{Bolz:2000fu} to be 
\be
\Gamma_q = \frac{\alpha\, g^2}{12}T^3\(\log\(\frac{T^2}{m_\gamma^2}\)+0.8194\), 
\ee
where $m_\gamma= eT/3$ is the plasmon mass in an electron-positron plasma and $T$ the temperature.
Taking the rate to be proportional to the number density of electrons,
$n_e=3\zeta(3)T^3/\pi^2$, we can generalise to a multicomponent plasma, 
\be
\Gamma_q \simeq \frac{\alpha\, g^2\, \pi^2}{36\zeta(3)}\(\log
\(\frac{T^2}{m_\gamma^2}\)+0.82\) n_q,
\ee 
where $n_q$ is the effective number density of charged particles, $n_q\!=\!\sum_i
Q_i^2 n_i\!\equiv\!(\zeta(3)/\pi^2)g_q(T) T^3$, where $Q_i$ is the charge of $i$-th particle species, 
and the parameter $g_q(T)$ represents the effective number of relativistic charged degrees of freedom. 
Also the plasmon mass has to be corrected by a factor $m_\gamma\propto g_q^{1/2}$ when more charged species are present.

The Primakoff process becomes inefficient when $\Gamma_q$ becomes smaller than the Hubble expansion parameter. In a radiation dominated universe, this is $H= 1.66\sqrt{g_*(T)} T^2/m_{\rm Pl}$ where $m_{\rm Pl}$ is the Planck mass and $g_*(T)$ is the effective number of relativistic degrees of freedom, which like $g_q(T)$ is a smooth function of $T$. 
We follow the definitions and notations of~\cite{Kolb:1990vq} for all the usual cosmological quantities.
The $T$ dependences of $g_q$ and $g_*$ are taken from the recent analysis of~\cite{Wantz:2009it}.
The ALP bath decouples at a temperature very sensitive to $g$
\be
T_{\rm fo}\simeq \frac{11}{\alpha\,g^2\, m_{\rm Pl}}\frac{\sqrt{g_*}}{g_q}=
123 \frac{\sqrt{g_*}}{g_q} 
\(\frac{10^{-9}\, {\rm GeV}^{-1}}{g}\)^2 {\rm GeV}. 
\ee

For values of $g \lesssim 2 \times 10^{-9}$ GeV, ALP interactions freeze out at temperatures above the electroweak scale ($E_{\rm EW}\sim 250$ GeV), where 
the particle content of the plasma is relatively speculative. 
For instance in the MSSM scenario, above the SUSY breaking energy scale we have
$g_*=228.75$, while the SM alone provides only 106.75 relativistic degrees of freedom.
We warn the reader that for $g \lesssim  10^{-17}$ GeV we require a 
most likely meaningless freeze out temperature above the Planck scale ($m_{\rm Pl}=1.22\times 10^{19}$ GeV).
We drew two dotted black lines for $T_{\rm fo}=E_{\rm EW}$ and for $T_{\rm fo}=m_{\rm Pl}$ in Figs.~\ref{fig:bounds} and~\ref{fig:boundsLife}.
 
After the freeze out, the number density of ALPs per comoving volume $N_\phi$
is conserved. Also its ratio with the comoving entropy density $Y_\phi=N_{\phi}/S$ is conserved.
Therefore at any later time (before the ALP decay), the ALP density will be given by
\be
\label{ALPyield}
n_\phi(T) = Y_\phi(T_{\rm fo})s(T) .
\ee
Given the standard particle content that we are assuming, there is a minimum ALP yield given by the value of 
$g_{*}(T_{\rm fo}>{\rm EW\ scale})$
\be \label{nphionngamma}
\frac{n_\phi}{n_\gamma}\geq \frac{1}{2}\frac{g_{*}(T)}{106.75}\simeq 0.005 g_{*}(T).
\ee

Cosmologically stable ALPs must not exceed the measured abundance of DM, therefore
\be
\Omega_\phi h^2 =\frac{\rho_\phi}{\rho_c}h^2=\frac{m_\phi\,n_{\phi}}{\rho_c}h^2<\Omega_{\rm DM}h^2=0.11
\ee
where $\rho_c=3 H^2m^2_{\rm Pl}/8\pi$ is the critical density and $h^2$ is the normalised Hubble expansion rate. ALPs with a mass $m_\phi=154$ eV would account for all the dark matter of the universe but larger masses are excluded. This exclusion is depicted in orange in our summary plots Figs.~\ref{fig:bounds} and~\ref{fig:boundsLife} and labelled as DM. The discovery of new degrees of freedom (dof) above the EW scale would relax this bound, which is linearly sensitive to $g_*(T_{\rm fo})$, by a factor $(106.75+\, {\rm new\ dof })/106.75$. This means that $\mathcal{O}(100)$ of them are needed for a sizeable change.

Let us now consider the dynamics of the ALP decay. 
The guiding quantity is the ratio of the decay rate over the Hubble parameter evaluated at $3T\sim m_\phi$ (when ALPs would become nonrelativistic if in thermal equilibrium) disregarding the ALP contribution
\be
\label{inversedecaycondition}
\frac{\decay}{H(m_\phi/3)}\simeq \frac{3.3}{\sqrt{g_*(m_\phi/3)}}\frac{m_\phi}{\rm MeV} \(\frac{g}{10^{-7}\, {\rm GeV}^{-1}} \)^2 .
\ee
If this ratio is large, the condition $\decay/H\sim 1$ is satisfied at a temperature higher than $m_\phi/3$, since $H$ grows with $T$. In these conditions, the inverse decay process, $\gamma\gamma\to \phi$ (photon coalescence) is effective and the ALP population regains thermal equilibrium~\cite{Cadamuro:2010cz}. 
Note that during this rethermalization process, a small amount of entropy will be generated. 
The population of ALPs will be kept in thermal equilibrium with photons by decays and inverse decays. 
When the temperature decreases to $T\sim m_\phi/3$, the thermal abundance starts to become Boltzmann suppressed and axions disappear from the bath in thermal equilibrium, very much like electrons and positrons annihilate at $T\sim m_e/3$. 

On the other hand, if the ratio is small, $\decay/H\sim 1 $ will happen when $T \ll m_\phi$ and ALPs are very nonrelativistic. 
In this case, inverse decays are Boltzmann suppressed and the ALP decay is not perturbed by the thermal bath of standard particles. 
It is also possible that the energy density in ALPs, $m_\phi n_\phi$, becomes larger than that of standard model particles, $(\pi^2/30) g_*(T)T^4$, 
and the ALP decays create a considerable amount of entropy. This happens if the decay temperature $T_{\rm d}$ satisfies
\be
\frac{m_\phi}{T_{d}} \gtrsim \frac{\pi^4}{30 \zeta(3)} \frac{g_*(T_{\rm fo})}{g_*(T_{\rm d})}\simeq 2.7 \frac{g_*(T_{\rm fo})}{g_*(T_{\rm d})} .
\ee

If ALPs do not dominate the universe energy budget when they decay $T_{\rm d}$ 
can be estimated by $\decay\sim H(T_{\rm d})$, which gives
\be
\label{decayT}
T_{\rm d} \simeq 
\frac{0.6}{g^{1/4}_*(T_{\rm d})}\(\frac{g}{10^{-7}\, {\rm GeV}^{-1}}\)
\(\frac{m_\phi}{\rm  MeV}\)^{3/2} {\rm MeV} . 
\ee
If the ALP energy density does dominate, the decay temperature is instead
\be
\label{decayT2}
T_{\rm d} \simeq 
\frac{0.7}{(g_*(T_{\rm d})/g_*(T_{\rm fo}))^{1/3}}\(\frac{g}{10^{-7}\, {\rm GeV}^{-1}}\)^{4/3}
\(\frac{m_\phi}{\rm  MeV}\)^{5/3} {\rm MeV} , 
\ee 
which is typically larger than the previous case (a matter dominated universe expands more slowly
than a radiation dominated one). If the universe becomes radiation dominated after the decay, the temperature in Eq.~\ref{decayT} gives the 
correct order of magnitude for the reheating temperature.
We evaluate the observable consequences of early ALP decay in the next sections.

\section{Early ALP decays} \label{sec:early}
\subsection{Neutrino dilution}

The ALP population decay produces photons whose energy and entropy tends to be
reshuffled between the different species that are present 
and active in the thermal bath at that moment. The relevant temperatures in the context of this paper are around and below the neutrino 
decoupling temperature,  $T\sim$ a few MeV. The active species to consider are then photons, electrons and neutrinos. 
Photons thermalize very fast with electrons, the particles most tightly coupled to neutrinos. Weak reactions between electrons and neutrinos heat the neutrino bath to keep track with the electron/photon temperature changes. 
The most relevant of these energy redistribution processes is $e^+ e^-\to \bar \nu \nu$. 
Since scattering processes such as $e^\pm \nu \to e^\pm \nu$ cannot change the neutrino number and are less effective in transferring energy to the neutrino bath, we neglect them\footnote{
Assume that electrons have a larger temperature. Then the speed of energy transfer per unit volume in scattering processes will be $\propto \langle \delta E \rangle T_e^4 T_\nu^4$ with $\langle \delta E \rangle$ a thermal-averaged energy transfer per scattering ($\sim T_e-T_\nu$); while for annihilations it will be $\sim T_e(T_\nu^8-T_e^8)$, which is much more sensitive to nonequilibrium situations.}. 
Therefore, if ALP decay happens after the freeze-out of the $e^+e^- \leftrightarrow \bar{\nu} \nu$ reactions, the electromagnetic energy and the entropy would not be shared with the neutrino bath, which then appears to have less energy than in the standard case.
The energy flow from ALPs to neutrinos can then be modelled by a set of Boltzmann equations for comoving energies, defined as
\be
X_i=\rho_i R^4  ,
\ee
where $R$ is the cosmic scale factor\footnote{$R$ has dimension of [energy]$^{-1}$ and so the $X$'s are dimensionless.}, and the ALP phase space distribution function $f(\tilde{k}_\phi)$ as a function of the comoving momentum, $\tilde{k}_\phi=R k_\phi$,
\begin{subequations}\label{eq:boltzmann}
\begin{align}
&\frac{d}{d\,t} f(\tilde{k}_\phi)=-(C_\gamma+C_q)(f-f^{\rm eq}) ,\label{eq:boltzgALP}\\ \nonumber
&\frac{d}{d\,t}(X_\gamma+X_e)= 3 H \delta p_e R^4 + \int \!\!\frac{d^3\tilde{k}_\phi}{(2\pi)^3} \tilde{\omega}_\phi (C_\gamma+C_q)(f-f^{\rm eq})\label{eq:boltzEM}\\
&\quad\quad\quad\quad\quad  +\frac{\Gamma_{e\nu}}{R^4} \(C_e\(X_{\nu_e}^2- {X_{\nu_e}^{\rm eq}}^2\)+C_{\mu\tau}\(X_{\nu_{\mu\tau}}^2-{X_{\nu_{\mu\tau}}^{\rm eq}}^2\)\)\; ,\\
&\frac{d}{d\,t}X_{\nu_e}=-\frac{\Gamma_{e\nu}}{R^4}  C_e\(X_{\nu_e}^2-{X_{\nu_e}^{\rm eq}}^2\) \; ,\\
&\frac{d}{d\,t}X_{\nu_{\mu\tau}}=-\frac{\Gamma_{e\nu}}{R^4} C_{\mu\tau}\(X_{\nu_{\mu\tau}}^2-{X_{\nu_{\mu\tau}}^{\rm eq}}^2\)\; ,\\
&\frac{d\, }{d\,t} R= \frac{1}{R}\sqrt{\frac{8\pi}{3 m_{\rm Pl}^2}\(X_{\gamma}+X_e+X_{\nu_e}+X_{\nu_{\mu\tau}}+\rho_\phi R^4\)}\; , 
\end{align}
\end{subequations}
where $\omega_\phi=\sqrt{k_\phi^2+m^2}$ is the ALP energy and $\tilde{\omega}_\phi=\omega_\phi R $. 
The collision terms for the decay and Primakoff processes are~\cite{Cadamuro:2010cz}
\begin{align}
C_\gamma&=\frac{m_\phi^2-4m_\gamma^2}{m_\phi^2}\frac{m_\phi}{\omega_\phi}
\[1+   \frac{2T}{k_\phi}\log
\frac{1-e^{-(\omega_\phi+k_\phi)/2T}}{1-e^{-(\omega_\phi-k_\phi)/2T}}\]\,
\decay,\\
C_q&\sim \frac{\alpha \, g^2}{16}n_e \log\[1+\frac{\(4\omega_\phi(m_e+3T)\)^2}{m_\gamma^2\(m_e^2+(m_e+3T)^2\)}\] . 
\end{align}

The equations~\ref{eq:boltzmann} describe the evolution of the comoving energy density stored in
$\gamma$ together with $e^\pm$, the three species of $\nu$'s, the ALPs and the cosmic scale factor $R$. 
The electron--neutrino energy exchange rate depends on the neutrino flavour because of the absence of charge current interactions for $\mu$ and $\tau$ flavours. At the low temperatures of interest
$\Gamma_{e\nu}\equiv G_{F}^2T_{\gamma}$ with $G_{F}$ the Fermi constant, and
$C_e\simeq 0.68$ and $C_{\mu\tau}\simeq 0.15$ which follows from the appropriated thermally averaged cross section. 
The factor $\delta p_e=p_e-\rho_e/3$ ($p_e$ and $\rho_e$ are the pressure and energy density of $e^\pm$) accounts for the comoving energy density
gain as electrons become increasingly nonrelativistic.
We assume that neutrinos always have a thermal distribution, determined only by
an effective temperature, which should be a reasonable first 
approximation and accurate enough for our purposes. We also neglect the energy reshuffling between 
different neutrino species, which does not influence the total neutrino density at leading order. 
The initial conditions are specified at $T\gg {\rm MeV}$ by having 
all species at a common temperature and the ALP number density given by
Eq.~\ref{ALPyield}. 

For values $g\lesssim10^{-7} {\rm GeV}^{-1}$, the Primakoff process is decoupled in the temperature range of interest and can be neglected. 
If the inverse decay is also negligible, we can integrate the ALP phase space distribution explicitly and, instead of using Eq.~\ref{eq:boltzgALP}, 
directly compute the evolution of the number density
\begin{equation}
\frac{d}{d\,t}\(n_\phi R^3\)=-\decay n_\phi R^3 ,  
\end{equation}
recovering the exponential decay law $N_\phi \propto  e^{-\decay t}$. The integral in Eq.~\ref{eq:boltzEM} is then 
\begin{equation}\label{eq:boltzmann2}
\int\frac{d^3\tilde{k}_\phi}{(2\pi)^3} \tilde{\omega}_\phi (C_\gamma+C_q)(f-f^{\rm eq})\simeq 
m_\phi \decay n_\phi R^4 .
\end{equation}

The final neutrino energy density is usually quoted as the effective number of standard neutrinos
\be
N_{\rm eff} = \frac{X_{\nu_e}+X_{\nu_{\mu\tau}}}{\frac{7}{8}\(\frac{4}{11}\)^{4/3}X_\gamma} . 
\ee 

\begin{figure}[tbp] 
   \centering
   \psfrag{g}{$g$}
   \subfigure[]{\includegraphics[width=3in]{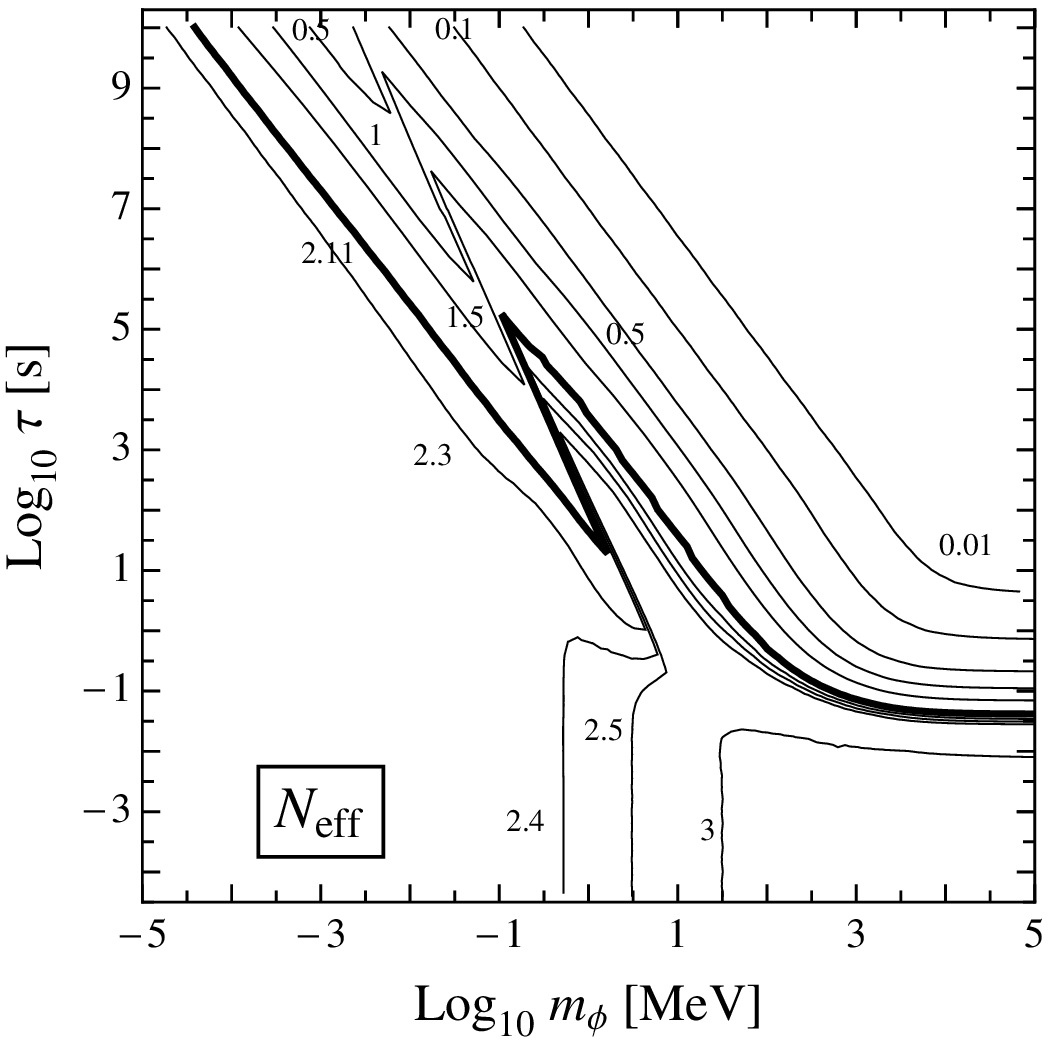}\label{fig:NeffA}}
   \subfigure[]{\includegraphics[width=3in]{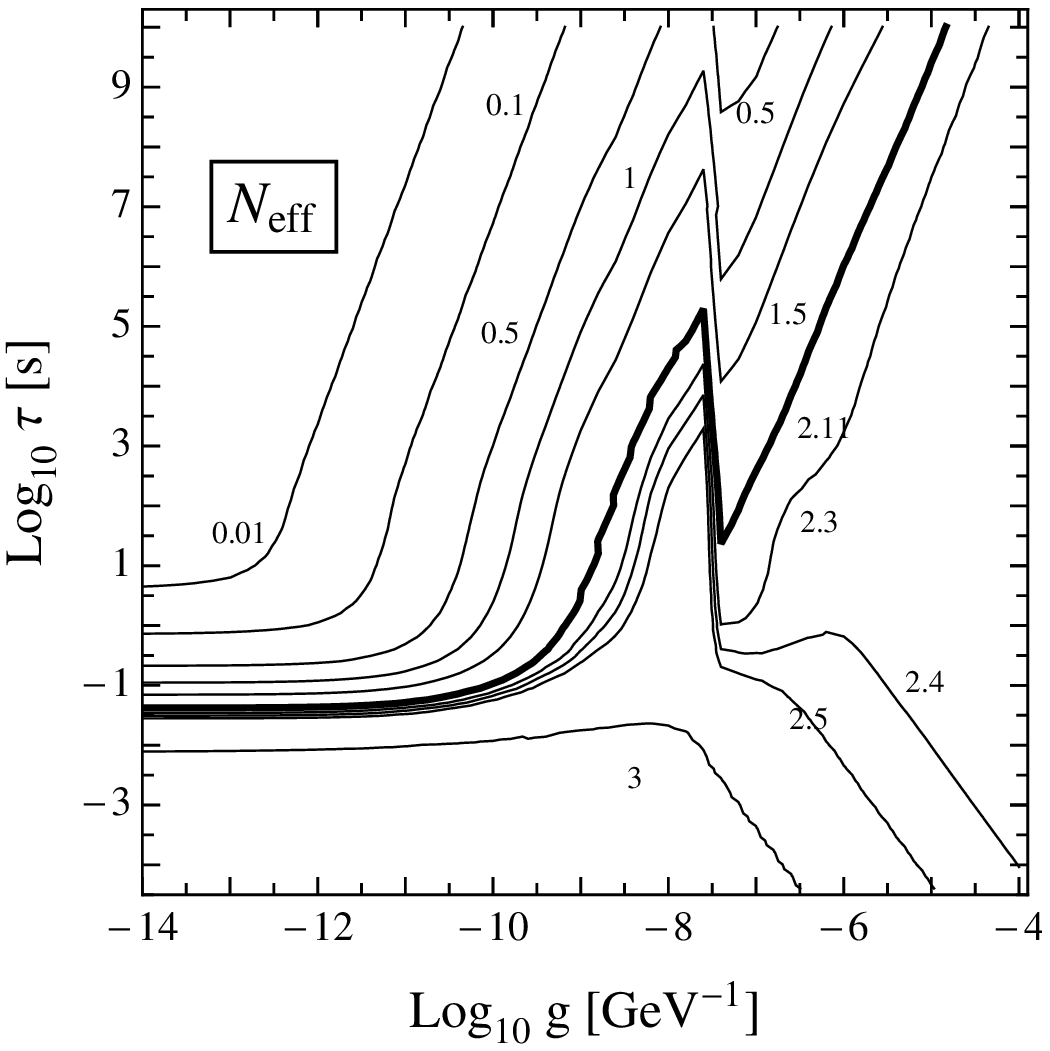}\label{fig:NeffB}}
   \caption{Contour plots of the number of effective neutrinos $N_{\rm eff}$ as a function of the ALP mass and lifetime (left) and of the coupling parameter $g$ and lifetime $\tau$ (right). 
   ALP cosmologies leading to $N_{\rm eff}<2.11$ can be safely excluded.}
   \label{fig:Neff}
\end{figure}

\begin{figure}[tbp]
\begin{center}
\psfrag{g}{\small $g$}
\psfrag{k}{{\scriptsize $\gamma$}}
\subfigure[]{\includegraphics[height=6.9cm]{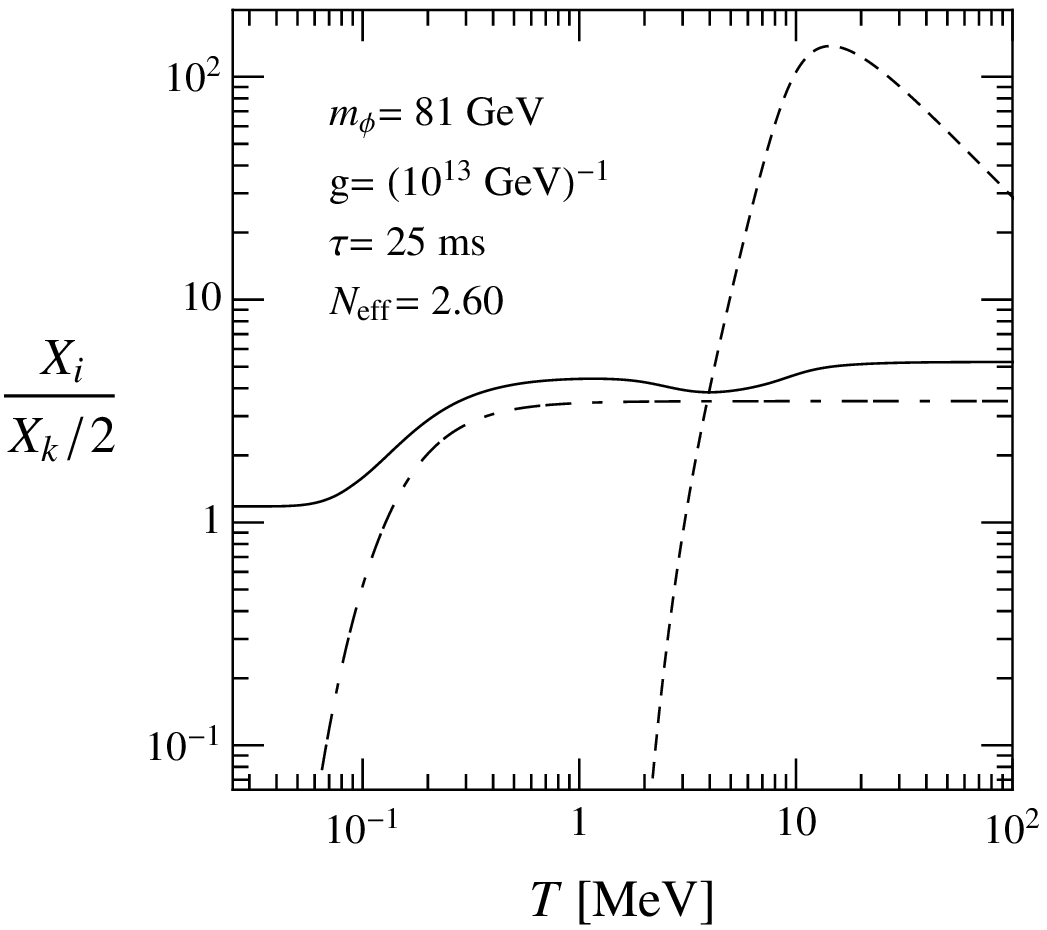} \label{fig2a}}
\subfigure[]{\includegraphics[height=6.9cm]{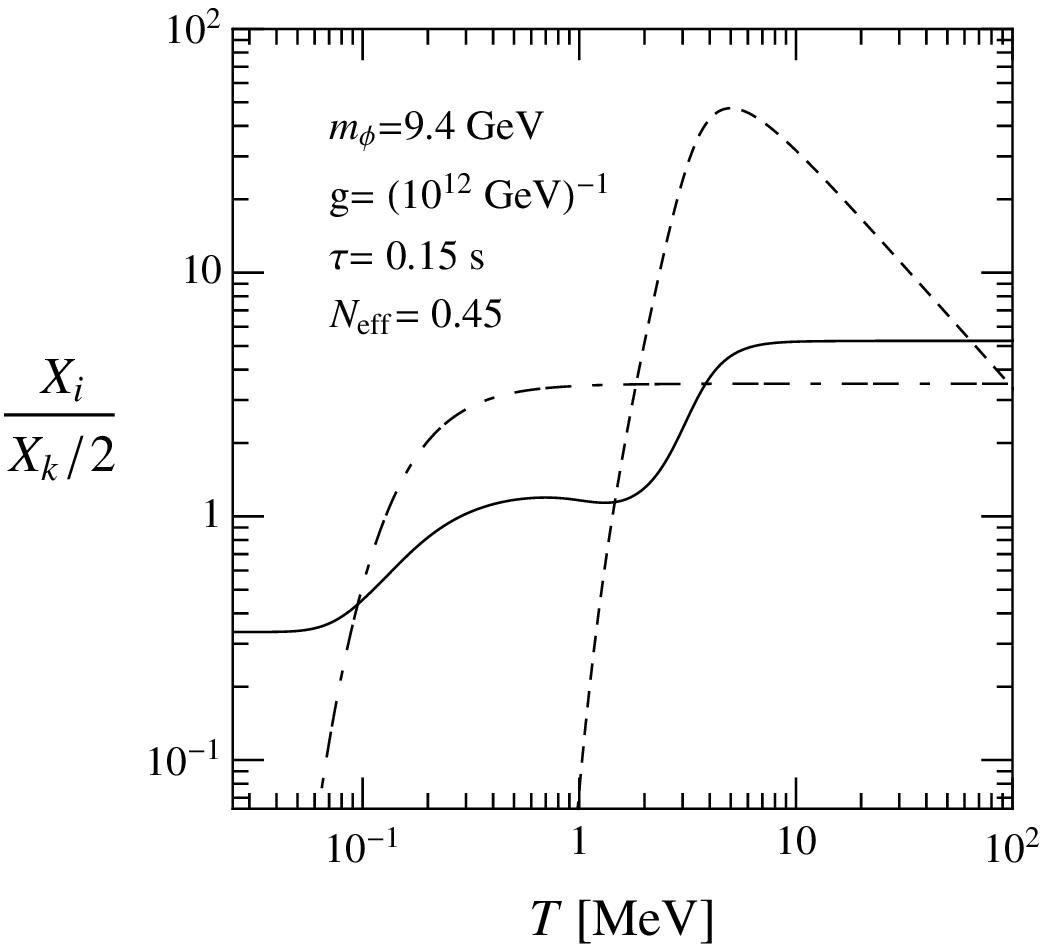} \label{fig2b}}

\subfigure[]{\includegraphics[height=7cm]{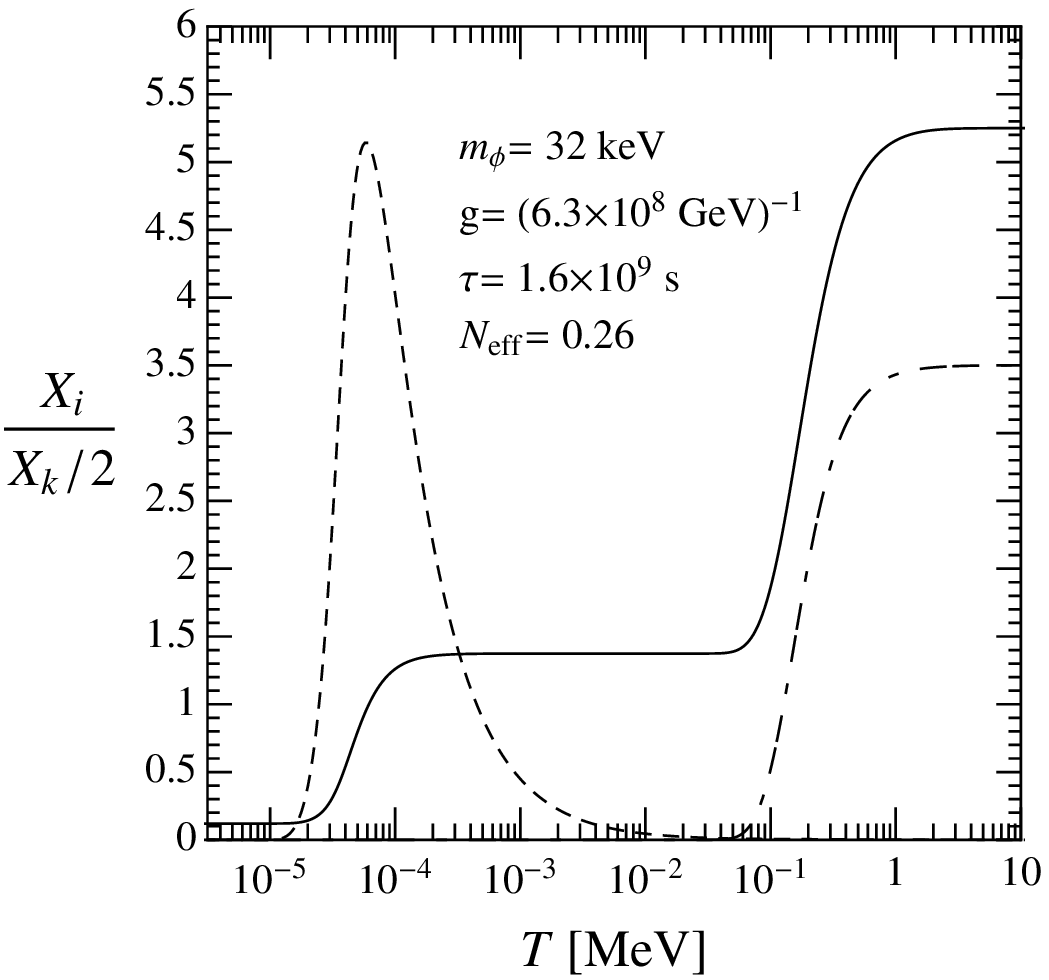} \label{fig3a}}
\subfigure[]{\includegraphics[height=7cm]{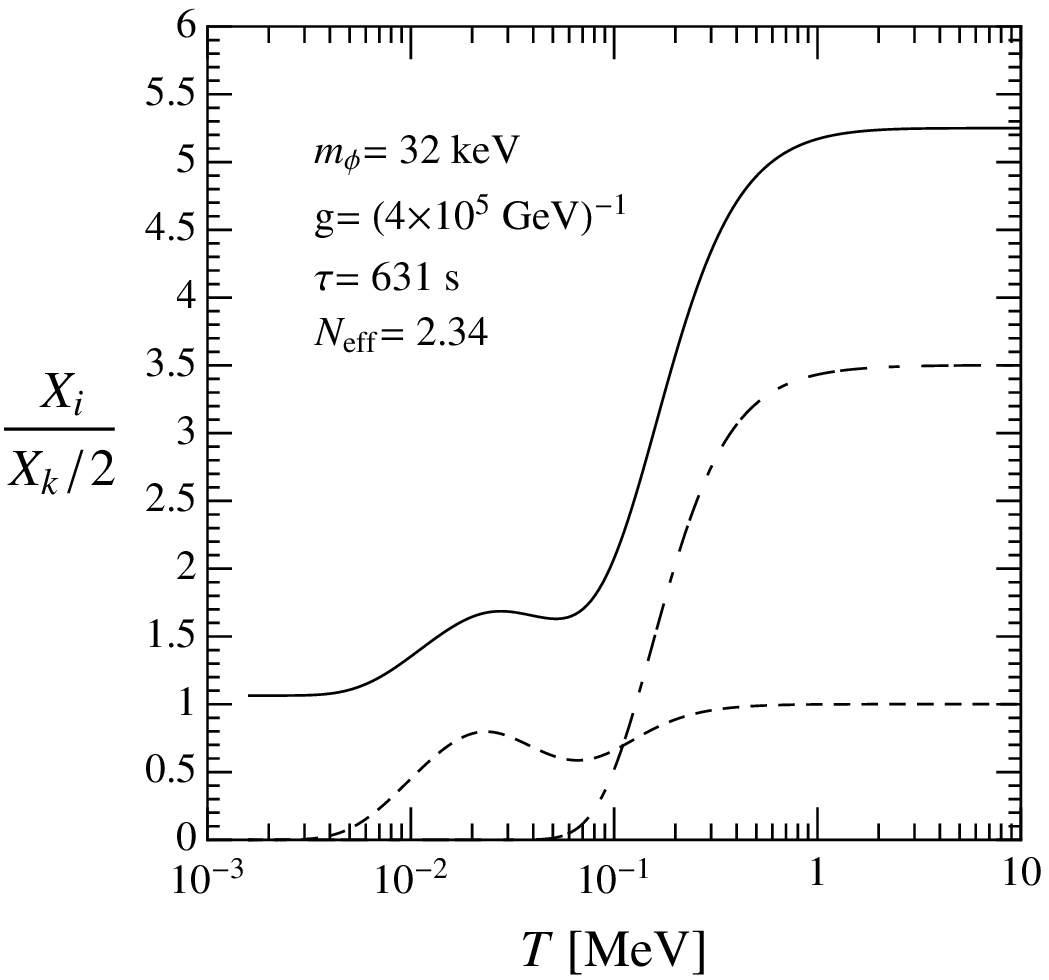} \label{fig3b}}
\end{center}
   \caption{Examples of the evolution of the comoving energy of all species of neutrinos (solid), ALPs (dashed) and electrons (double-dashed), as functions of the temperature of the universe. All energies are normalized to one thermal bosonic degree of freedom.}
   \label{fig2}
\end{figure}

We have scanned the ALP parameter space and present our results for $N_{\rm eff}$ in the $g-\tau$ and $m_\phi-\tau$ planes in Fig.~\ref{fig:Neff}. 
In Fig.~\ref{fig2}, we present some illustrative examples depicting the evolution of the $X'$s of electrons, neutrinos and ALPs as a function of the temperature. 
Note that in all of them, when ALPs become nonrelativistic, the ratio $X_\phi/X_\gamma$ rises 
because it becomes proportional to $m_{\phi}/T$, until the age of the universe becomes comparable with $\tau$. 

Fig.~\ref{fig2a} shows a typical case of very early decay of a massive ALP, when neutrinos are still partially coupled to electrons.
Here $N_{\rm eff}=2.6$, marginally different from the standard value of $3$.
Indeed, even if the ALP energy dominates the universe and the entropy injected during 
the decay is humongous, the reheating temperature is large enough for neutrinos to almost fully recover their thermal abundance. 
If the ALPs decay earlier the outcome is identical to standard cosmology since neutrinos will regain completely their thermal abundance.
In this and similar cases, the final value of $N_{\rm eff}$ is  related not to the total entropy, but only to that part injected after the freeze out of the neutrino-electron interactions. The neutrino dilution is sensitive  mainly to the ALP lifetime and not to 
$m_\phi$ or $g$ individually. This is precisely what we find in the lower right corner of Fig.~\ref{fig:NeffA}, and correspondingly to the lower left of Fig.~\ref{fig:NeffB}. 
Around $T\sim m_e$, electrons and positrons become nonrelativistic and annihilate, heating the photon bath (and baryons) but not the 
neutrinos, which have decoupled. The ratio $X_\nu/X_\gamma$ therefore decreases.
In this period, the photon temperature increases with respect to the neutrino temperature by the standard factor $(4/11)^{1/3}$ due to entropy conservation
\be
(T_ \gamma/T_\nu)^\prime=(4/11)^{1/3}(T_ \gamma/T_\nu).
\ee

The picture changes if we consider later decays. At a late enough point, the decays proceed when the  neutrinos have already decoupled; photons and electrons get all the ALP energy, as shown in Fig.~\ref{fig2b}.
The neutrino dilution is computable in this case. The ratio of the final and initial comoving entropies of the photon/electron bath  is~\cite{Kolb:1990vq}
\begin{equation}   \label{eq:entropy}
\frac{S_{f}}{S_{i}}=1.83\,{\langle g_{*S}^{1/3}\rangle}^{3/4}
\frac{m Y_{\phi}}{\sqrt{m_{\rm Pl}\decay}}. 
\end{equation}
where $\langle g_{*S}^{1/3}\rangle$, the time average of $g_{*S}$ during the decay, is an $\cal O$(1) factor.
The temperature of the electromagnetic bath increases with respect to the neutrino one by a factor 
$(S_{f}/S_{i})^{1/3}$ making $N_{\rm eff}=3(S_{f}/S_{i})^{-4/3}$ because also the electron entropy ends up in photons. 
The neutrino energy density is therefore strongly diluted by the energy gain of photons plus electrons. Note that this is mainly a function of  $m_\phi/\sqrt{\decay}\propto (g\sqrt{m_\phi})^{-1}$, which produces the characteristic slope of the isocontours at long ALP lifetimes $\tau\propto m^{-2}$ for Fig.~\ref{fig:NeffA} and $\tau\propto g^{4}$ for Fig.~\ref{fig:NeffB}. 
For $g\lesssim 10^{-9} {\rm GeV}^{-1}$, this is the only dependence on $g$, since $Y_\phi$ is constant.
Another example of late ALP decay, but with a smaller mass, is shown in Fig.~\ref{fig3a}. 
In this case we observe first the $e^{\pm}$ annihilation,  which heats photons with respect to neutrinos. A sizeable neutrino dilution is observable after the ALP decay. 

The relic abundance of ALPs grows for bigger values of $g$. In particular, around $g\sim 10^{-7.5} {\rm GeV}^{-1}$ it suffers an abrupt increase due to the sizeable decrease of $g_*$ during the QCD confining phase transition. The isocontours in Fig.~\ref{fig:Neff} have sharp features in this $g$ range (in Fig.~\ref{fig:NeffA} the $g$ dependence is hidden in $\tau$). 
ALPs decoupling at smaller temperatures, i.e.~with bigger $g$ values, do not get their abundance diluted by the QCD degrees of freedom: being more abundant, they produce more entropy when they decay. 

Finally, in Fig.~\ref{fig3b} we show an example for which inverse decays are relevant.
As the temperature drops, we observe a first decrease of the ALP energy density due to the electrons heating the photon bath. 
The inverse decay channel opens around $T\sim 70$ keV and helps 
ALPs to regain equilibrium before disappearing at $T\sim m_\phi$. During rethermalization, the photon energy \emph{decreases}, which can be seen as a slight rise in $X_\nu/X_\gamma$ (neutrinos are decoupled by then).  
Neglecting the entropy gain in the mixture process, which is typically small, entropy conservation gives 
$N_{\rm eff}=3(11/13)^{4/3}\simeq 2.4$~\cite{Cadamuro:2010cz}. 
Due to this mechanism, in the small-mass and short-lifetime region of the parameter space the prediction is $N_{\rm eff}\sim 2.4$. 
If $m_\phi\gtrsim$ a few MeV, the disappearance from the thermal bath happens when neutrinos are still coupled, so $N_{\rm eff}$ approaches 3. 
Since the disappearance of ALPs in local thermal equilibrium is governed only by the ALP mass, the isocontours of $N_{\rm eff}$ exactly follow the isocontours of $m_\phi$.

From the observational point of view, the presence of neutrinos modifies structure formation and leaves a trace in the anisotropies of the CMB~\cite{Kolb:1990vq,Dodelson:2003ft}.
The two main effects are a shift in the redshift of matter-radiation equality and the decrease of the density contrast at small scales caused by the neutrino free-streaming~\cite{Hamann:2007pi,Hamann:2010pw,GonzalezGarcia:2010un,Hamann:2010bk}. 
In our previous paper, we derived constraints from WMAP7 and other cosmological 
data on $N_{\rm eff}$ assuming a flat prior $0<N_{\rm eff}<3$. This is the range 
allowed in our scenario provided that ALPs have decayed before matter-radiation equality and the standard analysis of CMB is unperturbed by the presence of ALPs. See~\cite{Cadamuro:2010cz} for details and references. We found 
\begin{equation}\label{eq:neffconstraint}
N_{\rm eff}>
\begin{cases}
{2.70}&\hbox{at 68\% C.L.}\\
{2.39}&\hbox{at 95\% C.L.}\\
{2.11}&\hbox{at 99\% C.L.}
\end{cases}
\end{equation}  
Our numerical calculations exclude the yellow region in Figs.~\ref{fig:bounds} and~\ref{fig:boundsLife} (labelled $N_{\rm eff}$) at 99\% C.L.. Remarkably, the 95\% C.L. is just below the value $N_{\rm eff}=2.4$, predicted when ALPs are in thermal contact through Primakoff and inverse decays. The region disfavoured at the 95\% C.L. is much bigger than at 99\% C.L. but is extremely sensitive to the exact value of the bound so we prefer to quote the 99\% C.L.. 
As noted in~\cite{Cadamuro:2010cz}, the Planck satellite will not improve this bound, if it measures $N_{\rm eff}=3$ with the predicted error bars. The strength of the WMAP7 bound relies on  cosmology seeming to prefer $N_{\rm eff}>3$~\cite{Komatsu:2010fb}. Of course, if the Planck satellite, which is presently taking data, finds convincing evidence for extra radiation, much in cosmology will have to be reconsidered besides our ALP limits.

The mass lower bound of the bound is given by the limit of validity of our assumptions, i.e. standard cosmology at temperatures below 
the standard matter-radiation equality $T\sim 1$ eV. From Eq.~\ref{decayT} we estimate this to be
\be
\label{MReq}
\frac{g}{{\rm GeV}^{-1}}\gtrsim 10^{-3}\(\frac{\rm eV}{m_\phi}\)^{3/2} . 
\ee

\subsection{Big bang nucleosynthesis}\label{sec:BBN}

The decays of relic particles can have a considerable effect on the abundances of light elements predicted by big bang nucleosynthesis (see \cite{Sarkar:1995dd,Iocco:2008va,Pospelov:2010hj}). The impact depends strongly on the ALP mass, in particular whether ALPs are heavy enough to induce electromagnetic or hadronic cascades. We discuss these cases separately. 

\subsubsection*{Small masses}

If the ALP mass is smaller than a few MeV, the decay products cannot induce nuclear reactions and their effect on BBN is only indirect. 
The injected photons (and perhaps a small amount of electron/positron pairs) dilute both the neutrino and baryon densities, which in turn control the effectiveness of the standard nuclear reactions happening during BBN. 
The impact then depends on whether the decays happen before or after BBN.  

For decays happening after BBN, the injected photons heat the bath, decreasing the baryon to photon ratio%
\footnote{The value of $\eta$ can be measured at two different epochs, during BBN (estimated by measuring the primordial abundances) and at the CMB epoch (as imprinted in the temperature angular anisotropies).
These two estimates agree quite well, $\eta_{10}^{\rm CMB} =6.23\pm 0.17$~\cite{Komatsu:2010fb}  and $5.1<\eta_{10}^{\rm BBN}< 6.5$ at 95\% C.L.~\cite{Nakamura:2010zzi}. This agreement can be readily used as a constraint on any entropy injection between the BBN and CMB epochs. However, we also want to consider here the cases in which the ALPs decay during BBN, at least partially, where the application of this constraint is not sufficiently clear. },  $\eta$. 
Therefore, the value of $\eta$ during BBN is  larger than the one measured much later, at the CMB epoch by WMAP7 and other large-scale-structure data $\eta_{10}^{\rm CMB} =6.23\pm 0.17$~\cite{Komatsu:2010fb} ($\eta_{10}=10^{10} \eta$). 
If BBN proceeds in such a low-photon-density environment, the deuterium photo-disintegration reaction is less effective, so that a significant amount of deuterium forms earlier, i.e. the deuterium \emph{bottleneck} opens up earlier. 
This allows the subsequent reactions (burning deuterium into heavier elements) to happen at higher baryon densities, where they are more effective. 
The outcome of this high-$\eta_{\rm BBN}$ scenario is clear: intermediate nuclei like D or $^3$He are more easily consumed, and the final abundance of heavier nuclei like Li increases. 

The Helium abundance is determined by the neutron abundance during BBN, since essentially all neutrons end up in Helium nuclei. Let us then briefly review the history of neutrons. At high temperatures $T\gg$ MeV, electroweak reactions keep protons and neutrons in nuclear equilibrium, which predicts $n_n/n_p=e^{-Q_{pn}/T}$ ($Q_{pn}\approx m_n-m_p=1.293$ MeV) for the neutron/proton ratio. At $T\sim$ MeV, the reactions freeze out and afterwards neutrons can only decay until the deuterium bottleneck opens up, BBN commences and they get trapped in $^4$He nuclei. 
In the late ALP decay scenario, the bottleneck opens earlier and thus less neutrons decay, enhancing the final  $^4$He yield. But this is not the only effect. The presence of ALPs makes the universe expand faster,  which has two additional implications:
a) during the freeze-out of the $p\leftrightarrow n$ conversion reactions it induces an earlier freeze out and therefore a larger $n$ abundance and b) it shortens the time between this freeze-out and BBN and therefore the amount of neutrons that decay. 
Thus three mechanisms are responsible for the enhancement of the $^4$He yield.  

Let us now consider the case where ALPs decay \emph{before} BBN, i.e.~before the opening of the deuterium bottleneck. In this case $\eta_{\rm BBN}=\eta_{\rm CMB}$, and the main trends mentioned before disappear. 
However, the ALP decays can still modify BBN indirectly, if they happen between the freeze-out of weak interactions and BBN. There are three effects that we should take into account. First, ALPs are present during the freeze-out of the $p\leftrightarrow n$ reactions, so the $n$ abundance is in principle larger. Second, when ALPs decay they reduce $N_{\rm eff}$, as shown in the previous section. 
After this event, the universe expands slower than usual and therefore: a) more neutrons decay before BBN and b) BBN happens in a low $N_{\rm eff}$. 

As in the case of a post-BBN ALP decay, this scenario also implies low D and $^3$He and high Li, even if for a completely different reason (low $N_{\rm eff}^{\rm BBN}$ instead of high $\eta_{\rm BBN}$). 
The neutron concentration is however affected in two \emph{opposite} ways: higher initial $n$ abundance and post-decay \emph{low} $N_{\rm eff}$ giving neutrons more time to decay. These effects tend to compensate each other. The time that neutrons have to decay depends of how close to BBN the decays happen and indeed we find that $^4$He grows as the ALP decay happens closer to BBN and a small $^4$He region at $m_\phi\sim$ MeV and $\tau \sim 30$ s, where neutron decay plays a role inducing low $^4$He. However, because of the two opposing effects, the $^4$He abundance is not a sensitive indicator of ALPs in this region.

In order to numerically evaluate the impact of decaying ALPs in the BBN predictions, we have used a BBN code that includes the modified cosmology driven by ALP decays computed with the tools of the previous section. We have written a simple BBN code\footnote{When facing standard ALP-less cosmology, our results are in very good agreement with standard BBN calculations obtained with the KAWANO~\cite{Olive:1999ij} or \parte~\cite{Cuoco:2003cu} codes, given the theoretical and experimental uncertainties, which gives us confidence in our results. 
} in Mathematica to compute the primordial abundances of D, $^3$He, $^4$He, $^7$Li and $^7$Be. We have used the minimal reaction network relevant for $\eta\sim \eta_{\rm CMB}$ and $N_{\rm eff}\sim 3$ as detailed for instance in~\cite{Mukhanov:2003xs,Esmailzadeh:1990hf,Cuoco:2003cu}. 
This allows us to easily compute the outcome of BBN when ALPs have a non trivial role during BBN itself,  and gives the right trend when ALPs decay much later than BBN, enhancing enormously the value of $\eta_{\rm BBN}$ with respect to $\eta_{\rm CMB}$.  

\begin{figure}[tbp] 
   \centering
   \subfigure[]{\includegraphics[width=2.8in]{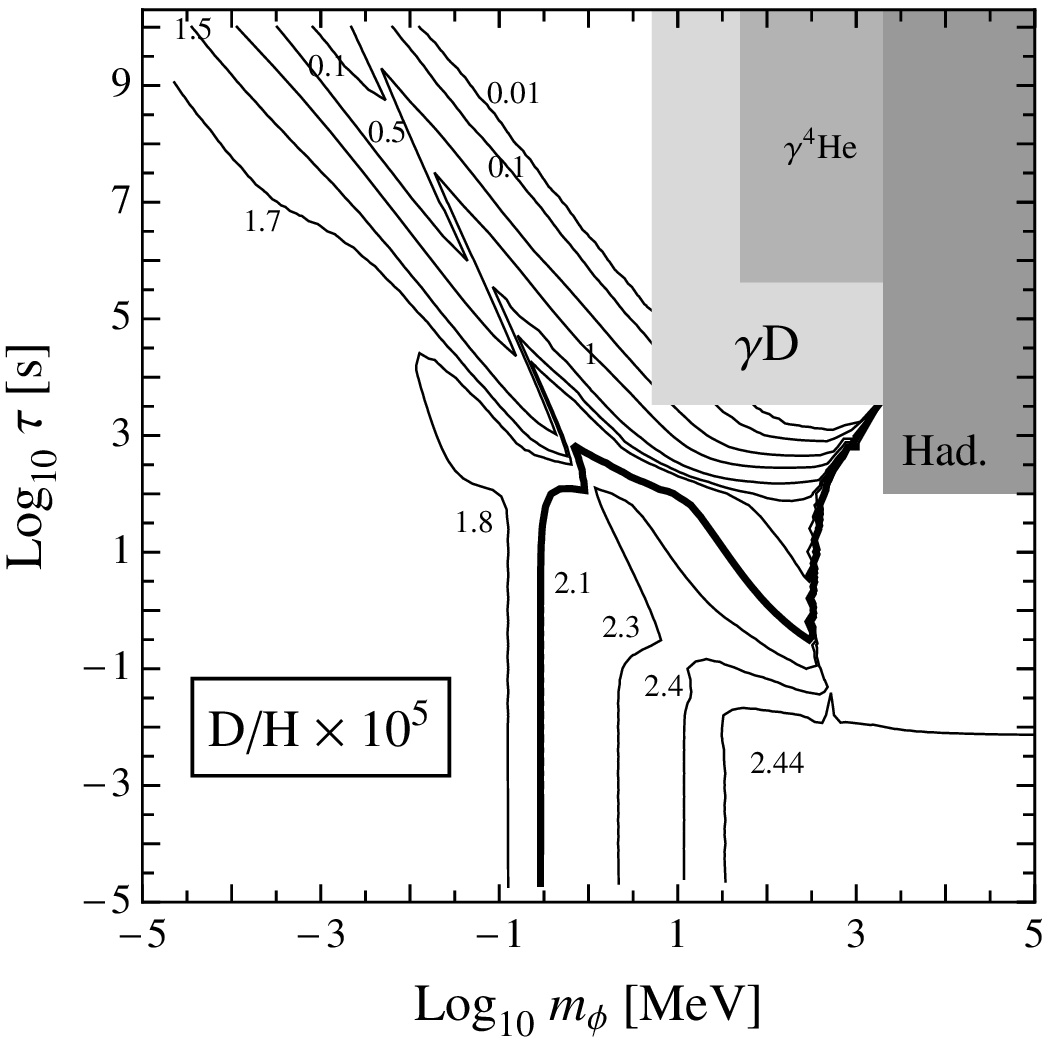}    \label{fig:D}}
   \subfigure[]{\includegraphics[width=2.8in]{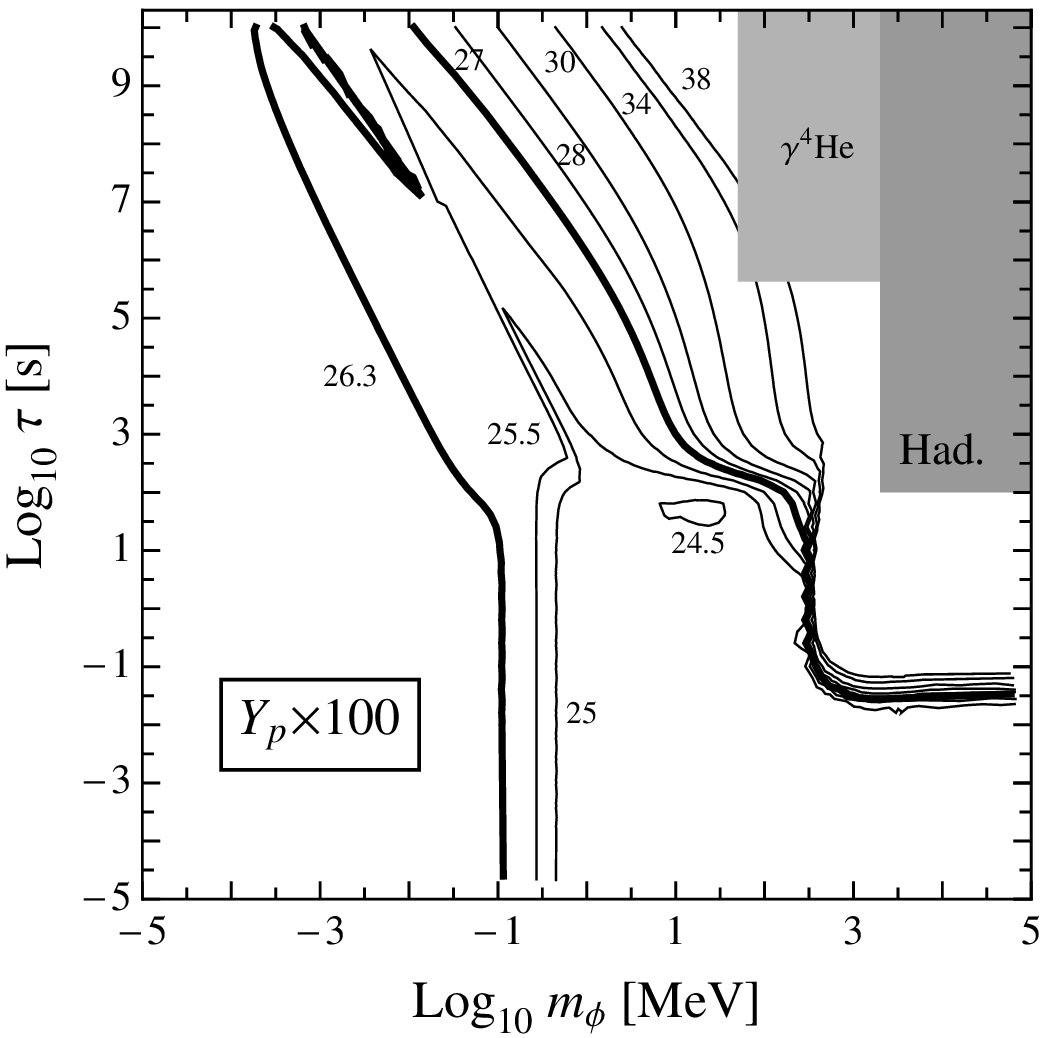}    \label{fig:He}}
   \caption{Isocontours of the primordial abundance of deuterium normalized to protons (D/H) (left) and helium $Y_p$ (right) in the decaying ALP cosmology, as a function of the ALP mass and lifetime.}
   \label{fig:DHe}
\end{figure}

The predictions for the primordial mass fraction of $^4$He, $Y_p=4 n_{\rm He}/n_B$, and the deuterium-over-proton ratio D/H in the ALP-decay scenario are shown in Fig.~\ref{fig:DHe}. The isocontours very much resemble those of $N_{\rm eff}$ because the outcome of BBN is mostly sensitive to the value of $\eta_{\rm BBN}$ and therefore to a possible baryon dilution, which qualitatively follows the same logic as the neutrino dilution.  
ALPs with small mass and fast decay disappear from the bath in LTE with photons at temperatures $\sim m_{\phi}$. Therefore BBN  only depends on the ALP mass, and not on their lifetime.  When the lifetime is longer, the isocontours are parallel to the lines of constant entropy production as in the $N_{\rm eff}$ case. We discuss the constraints at the end of this section. 

\subsubsection*{Large masses}

Photons from the decay of ALPs heavier than twice the photoionization threshold of elements can strongly modify the predictions of BBN. At high densities and temperatures, high-energy photons interact very fast with the thermal bath creating electromagnetic cascades, whose spectrum features a very sharp cut-off at high energies $E>E_C\sim m_e^2/(22 T)$~\cite{Kawasaki:1994sc}. 
At high $T$, this cut off will lay below the photo-dissociation threshold of nuclei and these effects are negligible. We have followed the methods of~\cite{Cyburt:2002uv} to check in which region of parameter space this phenomena can change our predictions. The photo-dissociation of deuterium ($Q=2.23$ MeV) will of course reduce its abundance further, making our bounds even stronger. 
On the other hand, the photo-dissociation of $^4$He can create deuterium, which reverses the trend of our constraints. 
This requires higher-energy photons since the Q-value is $19.81$ MeV and therefore a lower decay temperature.
We have depicted the areas where these effects are relevant as grey regions in Fig.~\ref{fig:DHe}. 

If $m_\phi$ is larger than twice the charged pion mass $m_{\pi^+}=139.57$ MeV, the decay channel $\phi\to \gamma \pi^+\pi^-$ opens up.  Even if its branching ratio is very small, the abundance of relic ALPs is almost thermal and thus a huge amount of pions (compared with that of the present nuclei) can be produced. 
If the decay happens before or during BBN, the universe is dense enough so that pions can induce neutron-proton interconversions $\pi^+ + n \to \pi^0+p^+$ and $\pi^- + p^+ \to \pi^0+n$ before decaying~\cite{Reno:1987qw}. The second reaction is favoured because of the Sommerfeld enhancement and the typical overabundance of protons over neutrons at $T<Q_{pn}$. 
These reactions will therefore tend to increase the neutron-to-proton ratio (they can do it much more drastically than the mere presence of the ALP during the $p\leftrightarrow n$ freeze out commented in the previous subsection).  
The higher neutron abundance would aid the heavy element production but it also increases D/H. 
This is the most important effect.  
Since almost all neutrons end up in $^4$He nuclei taking protons with them, a higher initial neutron abundance yields a smaller final proton abundance and thus a larger D/H ratio. 
If $n_n/n_p\simeq 1$ at the onset of BBN, all protons end up in $^4$He and D/H would be arbitrarily large!\footnote{Of course including neutron decay during BBN would still give a finite, albeit very large, result.}.

We have included the effects of pions in our BBN code following Ref.~\cite{Kawasaki:2004qu}.  
In Fig.~\ref{fig:DHe} we can see that the low D/H trend of low mass ALPs is drastically changed when the mass gets above $2m_{\pi^+}$ and the effect on $^4$He gets strongly boosted when crossing this boundary. The effects of pions are hampered if the ALPs decay very early ($\tau\lesssim 10^{-2}$ s), when the electroweak reactions $p^++e^-\leftrightarrow n +\nu_e$ can still re-establish the $n_n/n_p$ equilibrium;   or 
very late (above $\tau\sim 10^{2}$ s), because pions fail to interact before decaying\footnote{In this last tiny region our results cannot be  taken quantitatively on trust, since we have not taken into account the possibly ineffective slowing down of pions after $e^+e^-$ annihilation.}.
 
For ALP masses above a few GeV,  ALP decays will produce quark-antiquark pairs that will hadronize. Hadronic cascades can dissociate nuclei, if they are happening after BBN (typical time of $\tau\sim10^2$ s)~\cite{Cyburt:2009pg,Kawasaki:2004qu,Kawasaki:2004yh}. 
Because of the large ALP relic abundance, the effects of electromagnetic or hadronic cascades are necessarily dramatic.  
We consider extremely unlikely that the combined effect of nonstandard BBN with the post-BBN processing gives similar results to standard BBN. So we exclude all the regions where electromagnetic and hadronic cascades play a role, see Fig.~\ref{fig:DHe}.  
In the literature, cascade constraints are usually presented in terms of $m_{\phi}n_\phi/n_\gamma$, plotted in function of lifetime $\tau$. 
In the case in exam, it is possible to refer to this representation almost directly from Fig.~\ref{fig:DHe}, considering that the ratio $n_\phi/n_\gamma$, given by Eq.~\ref{nphionngamma}, is constant in all the plotted parameter space in which cascades play a role and does not change much outside.

\subsubsection*{Constraints}

\begin{figure}[tbp]
\begin{center}
\includegraphics[width=9cm]{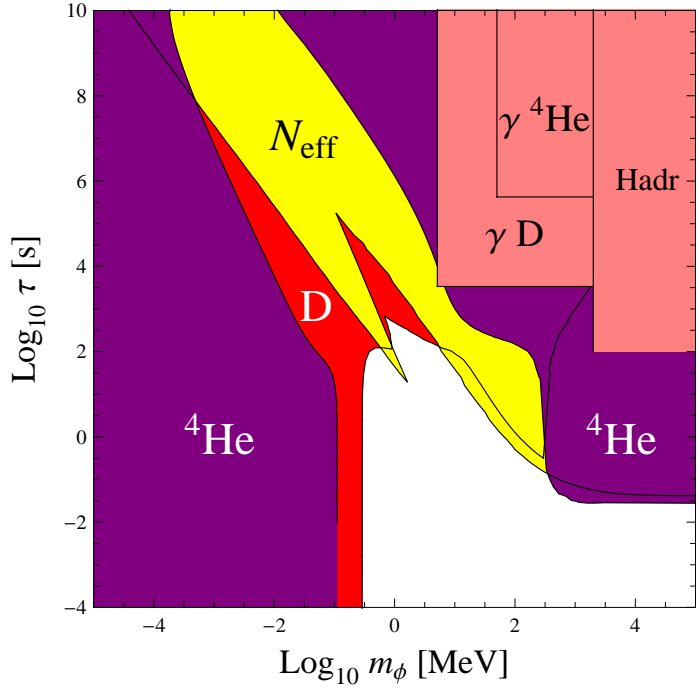}
\end{center}
   \caption{Bounds on early ALP decays from deuterium underproduction (D, red), helium overproduction (He, purple), 
helium photodissociation ($\gamma$He, pink), hadronic cascades (Hadr, pink) and neutrino dilution ($N_{\rm eff}$, yellow).}
   \label{summa}
\end{figure}

We can now establish constraints by comparing our predictions with the measurements of primordial abundances present in the literature. $^4$He was in the past the favourite indicator of the presence of extra degrees of freedom in the very early stages of BBN. The value of $Y_p$ can be estimated from an extrapolation to zero metallicity of the measured $^4$He content of metal-poor extragalactic HII regions. The systematics of the measurements, the extrapolation to zero metallicity together with a somewhat unknown early stellar nucleosynthesis have caused the best estimate of $Y_p$ to vary significantly over the years, a fact that calls for   extreme caution when quoting bounds. Here we shall be conservative and adhere to the proposal made in~\cite{Mangano:2011ar}, where the authors set a robust upper bound on $Y_p$ based on the assumption that the helium content is an increasing function of the metallicity of the cloud. They find
\be
\label{Yp}
Y_p < 0.2631 \quad (95\%\rm C.L.) . 
\ee
The corresponding exclusion is depicted as a purple region in Fig.~\ref{summa}, the summary plot of this section, and in Figs.~\ref{fig:bounds} and~\ref{fig:boundsLife}. 

When it comes to deuterium, we face other problems. The data are scarce, we only have reliable estimations from 7 high redshift low metallicity clouds absorbing the light of background quasars. The results of these estimations agree well at first glance but there is a scatter of the measurements beyond the expectations from the quoted systematics. 
The PDG quotes D/H$ =(2.82\pm0.21)\times 10^{-5}$ for the primordial ratio of deuterium to hydrogen, where the error has been enlarged to account for the still unexplained scatter. 
It is possible to obtain robust bounds from this measurement since the abundance of deuterium is known not to increase by the effect of unknown stellar processing. 
Any measurement of deuterium constitutes therefore a lower limit to the primordial value.  
In order to be conservative in this study we use 
\be
{\rm D/H}|_p >2.1\times 10^{-5} .  
\ee   
The corresponding exclusion bound is shown in red in Figs.~\ref{summa},~\ref{fig:bounds} and~\ref{fig:boundsLife}. 

Note that the PDG average agrees nicely with the outcome of BBN calculations with the value of $\eta_{\rm CMB}$ measured by WMAP, (we obtain D/H$|_p=(2.4\pm 0.1)\times 10^{5}$ where the error comes from the uncertainty in $\eta_{\rm CMB}$) providing one of the most beautiful tests of standard cosmology. 
A word of caution is however in order since the WMAP value depends on cosmological priors such as the spectral index of primordial fluctuations. 
The quoted value stems on a scale-free power-law which we carry as a further assumption.

The $^7$Li abundance cannot be reliably used to constrain ALP decays in the region of interest since at the moment observations do not agree with standard BBN predictions. The general trend is to increase the discrepancy between predictions and observations, since $^7$Li/H increases with $\eta_{\rm BBN}$, with a low $N_{\rm eff}$ and a high $n_n/n_p$ at BBN. This observation disfavours any reasonable attempt of solving the Li problem with decaying ALPs. Awaiting the Li problem to be solved, we have chosen not to propose any additional constraint although we observe that might theory and observations be reconciled, the sensitivity would be very similar to that of deuterium.

\subsection{Summary}

Our summary plot for bounds on early ALP decays is Fig.~\ref{summa}. 
Deuterium, helium or $N_{\rm eff}$ can be the most constraining argument depending on the ALP mass. Let us comment on why this is so. 
For very small lifetimes, ALPs will be in thermal equilibrium and the bounds only depend on the ALP mass.  
In this case deuterium dominates. 
The bound corresponds to ALPs that inject a a fraction of order 10\% of the total entropy in the electromagnetic bath after BBN. Under this circumstances the baryon and neutrino dilution is $\cal O$(1), and, since deuterium is the most sensitive observable to $\eta_{\rm BBN}$, it is also the most constraining argument. Note that the $^4$He abundance depends only logarithmically on $\eta_{\rm BBN}$ while D/H $\propto 1/(\eta_{\rm BBN})^{\sim 1.6}$. 
Moreover, we have already said that the $N_{\rm eff}$ bound is not enough to constrain the LTE decay of ALPs.

At  larger masses, ALPs in LTE would disappear before BBN and we loose the constraints for arbitrarily small lifetime. 
However, as the ALP lifetime increases, the decay proceeds increasingly out of equilibrium together with the entropy injection and the neutrino and baryon dilutions. 
For $m_\phi \lesssim$ 10 MeV we can only constrain decays happening during/after BBN ($\tau\gtrsim 10^2$ s) for which D/H is still the most constraining argument. 
But as the ALP mass increases, so it does the dilution and for $m_\phi \gtrsim$ 10 MeV it is big enough to be constrained by $N_{\rm eff}$. 
The bound on $\tau$ improves quadratically with $m_\phi$ (for a fixed entropy injection we have $m_\phi Y_\phi \sqrt{\tau}=$ const., 
c.f.~Eq.~\ref{eq:entropy}) until lifetimes of the order of the neutrino decoupling and then flattens. The deuterium bound follows quite closely behind 
$N_{\rm eff}$ because a low $N_{\rm eff}$ during BBN also implies a low D/H. Note that for lifetimes smaller than $10^2$ s, 
$\eta_{\rm BBN}=\eta_{\rm CMB}$ so the baryon dilution does not play any role and the low $N_{\rm eff}$ is the only responsible for the low D/H. 

For ALP masses above $2 m_{\rm \pi^+}$, the few pions produced in ALP decays affect notably the neutron/proton equilibrium before BBN leading to unacceptable high $^4$He. This effect dominates over the neutrino dilution because
 a very small amount of pions is needed to change $n_n/n_p$ while entropy injection must be ${\cal O}(1)$  to significantly affect $N_{\rm eff}$ (there are $10^9$ times less nucleons than photons!).
However, the $N_{\rm eff}$ bound, even relying on very different physics, lies very close. 
This can be easily understood. In this region the entropy injection is huge and the ALP rest mass dominates the energy budget of the universe at the decay. Nevertheless, the effects we are constraining are ${\cal O }(1)$ changes to standard cosmology. Therefore our constraints correspond to cases where most of the entropy release has been absorbed by thermalization processes, i.e.~most of the decays happen when the universe is strongly secured against non-thermal distortions. However, the very last ALPs, whose presence does not even affect the expansion significantly, are still able to produce an observable effect. The number of these ALPs depends exponentially on $\tau$ and therefore any bound on $\tau$ can only depend logarithmically upon any of the other quantities of the problem, such as the ALP mass, hadronic branching ratio, etc. 

Finally note that all bounds disappear for fast-decaying large-mass ALPs. These ALPs disappear from cosmology before the freeze-out of $p\leftrightarrow n$ reactions and neutrinos leaving no trace in the output of BBN. Here the ALP mass is crucial, because fast-decaying \emph{low mass} ($m_\phi \lesssim$MeV) ALPs are kept in equilibrium with photons through the inverse decay processes $\gamma\gamma\to \phi$ down to temperatures where they affect BBN, even if their lifetime would have made them decay much earlier. This is the reason why the BBN bounds do not disappear at low masses even for very short lifetimes in the figures of this section.
However, in this section we have assumed the decay of ALPs to occur before the CMB release epoch, in order to trust the standard analysis of CMB data that give us values for $\eta_{\rm CMB}$ and $N_{\rm eff}$. Therefore, we can not formally use our results to constrain ALPs which are kept in equilibrium longer than the time at which $N_{\rm eff}$ and $\eta$ are imprinted on the CMB. 
To be conservative, we take as a boundary time the matter-radiation equality, whose temperature is $T_{\rm eq}\sim 3$ eV, and thus our bounds cease at $m_\phi \sim 3 T_{\rm eq} \sim 10$ eV, see Fig.~\ref{fig:boundsLife}. 
Anyway, the existence of ALPs which decay in LTE and have a mass smaller than 10 eV is unconvincing from the cosmology point of view, as it would affect the CMB, for instance altering the transparency of the universe to photons. 
Besides this, such high coupling and light mass ALPs are definitely excluded by stellar evolution, positronium and $\Upsilon$ decays and helioscope searches~\cite{Jaeckel:2010ni}, which makes less important to know exactly where our bounds are no more valid.


\section{Late ALP decays}\label{sec:late}

We have just shown how it is possible to exclude some ALP parameter space if the decay happens before matter-radiation equivalence. Later decays can also be constrained, directly measuring the relic photons emitted or checking the distortion on the CMB spectrum they would create.

\subsection{Direct detection of ALP decays}

After recombination the universe becomes practically transparent to radiation,
since almost all the electrons are captured by nuclei forming neutral atoms.
The photons injected by ALP decay can be in principle directly detected, unless their wavelength lies in the ultraviolet range and they are absorbed in the photoionization process of atoms. 

In the parameter space we can constrain, the ALP decays happen at rest in the comoving frame. The spectral flux of photons produced in the decay of a diffuse ALP population is~\cite{Masso:1997ru,Masso:1999wj}
\begin{subequations}\label{eq:flux}
\begin{align}
\frac{dF_E}{dEd\Omega}&=\frac{1}{2\pi}\frac{\decay}{H(z)}\frac{n_{\phi}(z)}{(1+z)^3}=\\
&\simeq \frac{\bar{n}_{\phi 0}}{2\pi\tau H_0}\(\frac{E_0}{m_\phi/2}\)^{3/2}\exp\(-\frac{t_0}{\tau}\(\frac{E_0}{m_\phi/2}\)^{3/2}\),\label{eq:fluxMD}
\end{align}
\end{subequations}
where the subscript 0 means quantities at present time, $\bar{n}_{\phi 0}$ is the putative ALP number density if $\phi$ would be stable and $E_0$ is the energy at which the photon would be seen today. 
The photon initial energy is $m_\phi/2$ and thus the redshift of the decay is given by $1+z=(m_\phi/2)/E_0$, i.e.~the ratio between the emitted (then) and the measured (today) energy. 
For simplicity we assumed matter domination neglecting $\mathcal{O}(1)$ corrections due to the  cosmological constant. 

\begin{figure}[tbp]
\begin{center}
\includegraphics[height=9cm]{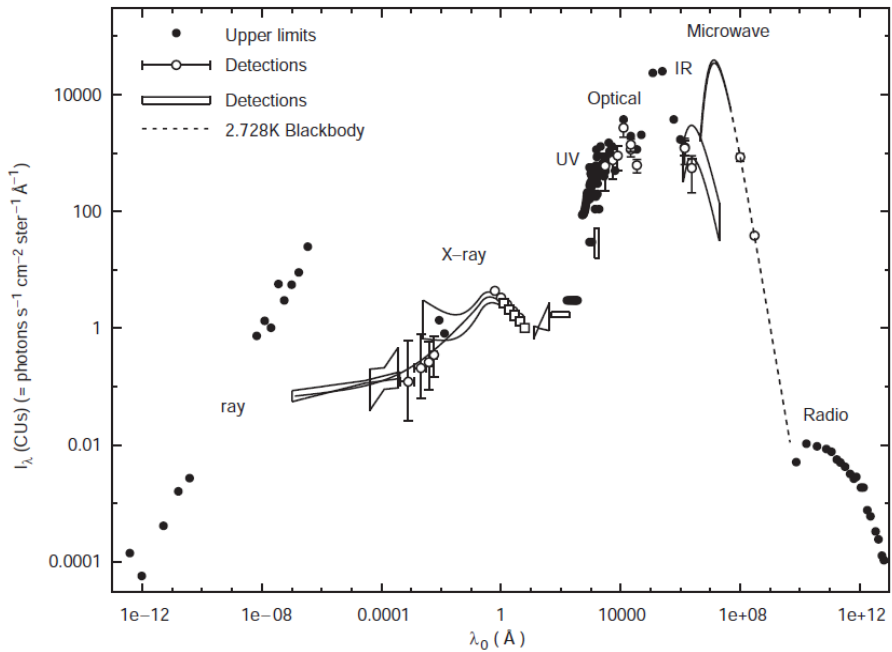}
\end{center}
   \caption{Extragalactic photon background~\cite{Overduin:2004sz}. Courtesy from J.~Overduin and P.~Wesson.}
   \label{fig:bkg}
\end{figure}

Radiation above $13.6$ eV can photoionize hydrogen, an effect that we take into account correcting the flux by multiplying Eq.~\ref{eq:flux} with the survival probability
\begin{align}
P(z)&=e^{-\kappa(z,E)}\\
\kappa(z,E)&=\int_0^z \frac{dz^\prime}{H(z^\prime)(1+z^\prime)}n_H(z^\prime)\sigma_{\rm pe}(E),
\end{align}
where $E=E_0(1+z)$, $n_H$ is the hydrogen number density and $\sigma_{pe}=256\pi (E_{1s}/E)^\frac{7}{2}/(3\alpha m_e^2)$ is the Hydrogen photoelectric cross-section with $E_{1s}=13.6\,{\rm eV}$.  
We have thus compared the resulting spectrum with the extragalactic background light (EBL) spectrum we have found in the review by Overduin and Wesson~\cite{Overduin:2004sz}, reproduced in Fig.~\ref{fig:bkg}. The region of parameter space excluded by this comparison is plotted in dull green in Figs.~\ref{fig:bounds} and~\ref{fig:boundsLife} and it is labelled EBL. The same approach was used in Refs.~\cite{Overduin:2004sz,Ressell:1991zv} to look for axions in the optical EBL. 
In this parameter space the ALP abundance would be the minimum considered $n_\phi/n_\gamma=(3.36/106.75)/2\simeq 0.016$. As we noted, for $m_\phi>154$ eV the ALP energy density overcloses the universe, in gross contradiction with observations. Therefore, above this mass we present our bounds \emph{assuming that ALPs provide the right amount of DM}, which of course requires some non-standard dilution of the ALP number density by additional degrees-of-freedom above the electroweak scale. 

The bound gets severely degraded in the $m_\phi=13.6-300$ eV range where not only absorption is very strong but also the 
experimental data are extremely challenging and the EBL spectrum has only an upper bound estimate (of course these facts are closely related).

A more powerful strategy to search for decay photons is to examine the light emitted in galaxies and large scale structure, where the dark matter density is above the average. 
Also the backgrounds are above the EBL, but the decay photons show up as a peak at frequency $\omega=m_\phi/2$ in the galactic spectra, which helps background subtraction. 
Searches in the visible have been presented in Refs.~\cite{Ressell:1991zv,Bershady:1990sw} and more recently in Ref.~\cite{Grin:2006aw}, and in X-rays in Refs.~\cite{Boyarsky:2006fg,Boyarsky:2009ix} searching for sterile neutrino dark matter decays. We followed these references, rescaling their results for the ALP decay mode and abundance (for $m_{\phi}<154$ eV), and we obtained the exclusion bounds in Figs.~\ref{fig:bounds} and~\ref{fig:boundsLife} labelled respectively Optical and X-Rays. 
We have also plotted in Fig.~\ref{fig:boundsLife} the exclusion bound obtained by galactic line searches in the $\gamma$-ray range for decaying DM, labelling it $\gamma$-rays. 
We took the data from~\cite{Vertongen:2011mu} and again we rescaled them for the ALP decay mode.

The radiation produced by ALP decay, being proportional to the
relic ALP density, would be linearly sensitive to the particle content above the EW scale through $g_*(T_{\rm fo})$. 
The constraint on $g$, for $m_{\phi}<154$ eV, relaxes as $\sqrt{1+{\rm new\, dof}/106.75}$ with the new thermal degrees of freedom with masses between the EW and $T_{\rm fo}$.

\subsection{Distortions of the CMB spectrum}

The constraint that allows to fill the gap between the BBN and $N_{\rm eff}$ excluded region and the telescope bounds (and also largely overlaps with these) is given by distortions of the CMB spectrum.
FIRAS measured the CMB spectrum to be a perfect black-body within the experimental errors~\cite{Fixsen:1998kq,firas}.
Any amount of electromagnetic radiation from the ALP decays should have had enough time to fully thermalize or must be very tiny compared with the CMB itself, ${\cal O}(10^{-5})$~\cite{Fixsen:1998kq}. 
The ALP energy injection is always greater than $10^{-5}$, and thus is potentially in radical conflict with observations, unless ALPs have not decayed significantly yet. On the other hand if the decay is very early, the distortions can be rethermalized.
Evading these arguments requires cosmological scenarios with $g_*(T_{\rm fo})\gtrsim {\cal O}(10^{5})$.
 
The most important processes for the thermalization of injected photons are double Compton scattering, $e\gamma\leftrightarrow e\gamma\gamma$, and Bremsstrahlung $p e\leftrightarrow p e \gamma$, which permit to fully regain the thermodynamical equilibrium distribution. In standard cosmology double Compton dominates\footnote{Actually in the decaying ALP cosmology this will be the case as well. Even if the value of $\eta$ before the decay can be much lager than $\eta_{\rm CMB}$, the amount of photons injected by ALPs is so huge that even the latest ALP decays are dangerous. The latest ALPs are not energetically significant and decay when $\eta\sim \eta_{\rm CMB}$ so double Compton will dominate in their thermalization.} because of the small value of $\eta$, being extremely effective until the temperature drops below $T_{\rm DC}\sim750$ eV.  
This leaves Compton scattering $e\gamma\to e\gamma$ as the most relevant interaction between photons and the rest of the plasma. 
Through Compton scattering only kinetic equilibrium can be obtained, as it can redistribute the energy of photons but not  change the photon number, thus leaving a non-zero degeneracy parameter~\cite{Hu:1992dc}
\be
\mu\simeq\left.\frac{1}{0.714}\(3\frac{\rho_\phi}{\rho_\gamma}-4\frac{2 n_\phi}{n_\gamma}\)\right|_{T=T_{\rm d}},
\ee
which is constrained by FIRAS to be $|\mu|<0.9\times 10^{-4}$~\cite{Fixsen:1998kq}.
ALPs with decoupling temperature smaller than $T_{\rm DC}$ are excluded while  
in the opposite case the produced distortions can be efficiently erased.
The light green region labelled CMB $\mu$ in Figs.~\ref{fig:bounds} and~\ref{fig:boundsLife} corresponds to $T_{\rm d}<T_{\rm DC}$. 
Note that this bound is somewhat conservative, as a significant amount of energy can be released after $T_{\rm d}$, especially above $m_\phi\sim$ keV where the ALP energy dominates the universe before the decay.

At $T_{\rm C}\sim25$ eV also Compton scattering freezes-out and the above bound has to be reformulated.
Electrons rapidly thermalize with the non-thermal population of photons, but the CMB can not drain efficiently energy out of them. The degree of thermalization that CMB photons can still attain depends on their energy and this imprints a typical pattern on the CMB spectrum. The parameter $y$, characterizing this distortion, can be estimated
as~\cite{Masso:1997ru}
\be
\left.\frac{\rho_\phi}{\rho_\gamma}\right|_{T=T_{\rm d}} \simeq \exp(4y)-1
\ee
and is bound to be $|y|<1.5\times 10^{-5}$~\cite{Fixsen:1998kq}. 
The region excluded by this last bound is roughly  $T_{\rm d}<T_{\rm C}$ and is shown in Figs.~\ref{fig:bounds} and~\ref{fig:boundsLife} coloured in light green and labelled CMB y.

\subsection{Reionization history}

At temperatures around $0.3$ eV (redshift $z\sim1300$), most electrons and protons combine into neutral Hydrogen and the universe becomes effectively transparent to the CMB. The universe reionizes again much later, between redshifts $6\sim10$ presumably due to ultraviolet emission from the first galaxies but the details of this process are still poorly understood. The details of the reionization process can be studied by the slight imprint that the free electrons leave in the CMB through Thomson scattering, for instance in the polarization.  
The optical depth for CMB photons is one of the parameter that can be measured from the CMB and it is defined as
\be
\tau_{\rm opt}(z_1,z_2)=-\int_{z_1}^{z_2}\frac{\sigma_T n_e(z) x_{\rm ion}(z)}{H(z)(1+z)}dz,
\ee
where $\sigma_T$ is the Thomson cross section, $n_e$ is the total electron density and $x_{\rm ion}$ the fraction of them not trapped in nuclei, i.e. the free electron or ionized fraction.
The WMAP7 measured $\tau_{\rm opt}$ after recombination to be $0.088\pm0.015$~\cite{Komatsu:2010fb}. 
A factor $0.04-0.05$ of this can be attributed to a fully ionized universe up to redshift $\sim 6$, as supported by the absence of Ly-$\alpha$ features in quasar spectra. 
The origin of the remaining fraction, $\tau_6\simeq 0.04$, is still uncertain.

We already commented that it seems hopeless to detect directly ALPs decaying into ultraviolet photons (energy range $13.6\sim 300$ eV), because they photoionize very efficiently Hydrogen and they are thus absorbed very fast. Of course, these ALPs contribute to the reionization story and they can be constrained indirectly, through their contribution to the CMB optical depth. 

It is simple to compute a rough estimate of the ionization fraction $x_{\rm ion}$ caused by the ALP decay photons~\cite{Natarajan:2010dc}. Assuming that each decay photon ionizes only one H (immediately after the ALP decay) the number of ionizations per unit time can be estimated as
\be
\xi(z)\sim 2 \decay \frac{ n_{\phi}(z)}{n_{H}(z)}
\sim 2 \times 10^{-3}\(\frac{m_\phi}{100\ {\rm eV}}\)^3\(\frac{g}{10^{-13}\ {\rm GeV}^{-1}}\)^2e^{-\frac{2}{3}\frac{\decay}{H(z)}}\ {\rm Myr}^{-1}.
\ee
If we now multiply this quantity with a typical time scale~\cite{Natarajan:2010dc}
\be 
t_H=1/H(z)\sim2.4\ {\rm Myr} \(501/(1+z)\)^{3/2}, 
\ee 
we get a conservative estimate of the typical degree of ionization induced by the ALP decays up to a certain redshift. ALPs with 100 eV mass and $g\sim10^{-13}$GeV$^{-1}$ would have produced an ionization comparable with the standard residual value $4\sim 10^{-4}$ already at high redshifts $\sim 500$, and the ionization grows in time as $(1+z)^{-3/2}$ showing a potentially interesting effect. Indeed one could even think that ALPs close to these parameters may have had a role in reionizing the universe, but the $(1+z)^{-3/2}$ dependence is too soft. Reionization seems to be a much more abrupt process. 

In order to obtain a more detailed constrain, we calculated the ionization history of the universe in the decaying ALP cosmology by introducing the ALP ionizations in the recombination code RECFAST~\cite{astro-ph/9909275}. Then, we computed the optical depth in the interval $z=6-100$, requiring it to not exceed $\tau_6$. In the calculation, we used the ALP thermal abundance and considered only a ionization for each decaying ALP, which is certainly conservative. 
Our results are excluding the light green region labelled $x_{\rm ion}$ in Figs.~\ref{fig:bounds} and~\ref{fig:boundsLife}. 
This bound would increase up to one order of magnitude at the largest masses for which ionization is effective, $m_\phi\lesssim 300$ eV, if we assume that all the energy of the emitted photons can be converted into ionization.



\section{Beyond the two photon coupling}\label{sec:beyond}

A generic pseudo Nambu-Goldstone boson can feature other couplings to SM particles~\cite{Gelmini:1982zz}: anomalous couplings to two gluons
\begin{equation}
{\cal L}\ni c_{gg} \frac{\phi}{f_\phi}\frac{\alpha_s}{4\pi}{\rm Tr}\{G_{\mu\nu}\widetilde G_{\mu\nu}\}
\end{equation}
and derivative couplings to fermions $f$, possibly even flavour non-diagonal
\begin{equation}
{\cal L}\ni  c_{ff'}\frac{\partial_\mu \phi}{2f_\phi}\bar f \gamma_5\gamma^\mu f'
\end{equation}
with, in principle, ${\cal O}(1)$ coefficients $c_{gg},c_{ff'}$. 
The addition of other couplings to SM particles necessarily implies a smaller temperature for the ALP decoupling and therefore a higher primordial abundance. 
In this sense, considering only the two-photon coupling for the relic ALP production produces conservative bounds. All what remains is the ALP decay, for which the new couplings above open new decay channels.  
The coupling to fermions allows the decay $\phi\to \bar f f' $ at a rate
\begin{equation}
\Gamma_{\phi\to \bar f f'} =
\( \frac{c_{ff'}}{f_\phi}\)^2  
\frac{\(m_f+m_{f'}\)^2m_\phi}{16\pi}
\sqrt{1-\(\frac{m_f+m_{f'}}{m_\phi}\)^2}
\(1-\(\frac{m_f-m_{f'}}{m_\phi}\)^2\)^{3/2} . 
\end{equation}
which is suppressed with respect to the two-photon decay for small fermion masses.
Writing $g\equiv c_{\gamma\gamma}\alpha/(2\pi f_\phi)$ the $\phi\to \bar f f'$ can dominate only in an interval near the kinematic threshold $1>(m_f+m_f')/m_\phi\gtrsim \alpha c_{\gamma\gamma}/4\pi c_{ff'}$.  
For ALP masses above few GeV, the coupling to gluons allows the ALP decay into two gluons at a rate $\Gamma_{\phi\to gg}= 8 (c_{gg}/c_{\gamma\gamma})^2 \Gamma_{\phi\to \gamma\gamma}$. At low masses this coupling implies a phenomenology very similar to the axion case\footnote{This particle cannot solve the strong-CP problem unless it is built-in massless or nearly massless. 
All our bounds apply to particles more massive than the axion so there is no point in considering this possibility here.}. The ALP will mix with the $\eta'$ and, through it, with the pseudoscalar mesons and hadrons and get new contributions to the two-photon coupling. 
 
Let us now review the impact of these new decay channels in our bounds.
First of all note that if the ALPs are cosmologically stable, the bounds from direct detection of ALP decay photons and the DM overproduction still hold. These arguments span the long lifetime range of our constrained region.
The short lifetime region corresponds to BBN and $N_{\rm eff}$ constrains and it is summarised in Fig.~\ref{summa} in the ALP mass and lifetime plane. 
At low masses, the deuterium and He bounds come from ALPs in thermal equilibrium with the bath. 
Clearly, adding more couplings between the ALP and SM particles we cannot avoid these bounds. 
In the intermediate mass region 300 keV $\lesssim m_\phi\lesssim $ 2 $m_\mu$, where $m_\mu=105.7$ MeV is the muon mass, the D/H and $N_{\rm eff}$ bounds follow from the dilution of baryons and neutrinos with respect to photons. 
These bounds apply to ALPs decaying into photons or electrons (actually we have not made a difference between the two in our equations) and since the direct decay into neutrinos is suppressed by $\sim (m_\nu/m_\phi)^2$, an amazingly tiny number\footnote{ Unless one considers sterile neutrinos with $m_\nu\sim m_\phi$ but then neutrinos have a strong tendency to constitute too much DM. A way to avoid this is to make them decay into SM neutrino + photon, but this produces entropy so we expect a similar, slightly smaller, bound from D/H in this case. In these models the low $N_{\rm eff}$ tendency is reversed since the sterile neutrinos produce neutrinos in its decay.}, the bounds are perfectly valid provided one interprets $\tau$ as the total lifetime (not only due to the two photon decay channel). If the decay into two electrons dominates, when we translate the bounds of Fig.~\ref{summa} in the $m_\phi-g$ plane they will show \emph{worst} than if we only consider the two photon coupling. The lower bound on $g$ reduces by a factor $\sim 2\pi c_{e e} m_e/\alpha c_{\gamma\gamma} m_\phi$. 

If $2 m_{\pi^+}>m_\phi>2 m_\mu$ we have a somehow different scenario where the ALP tends to favour the $\phi\to \mu^+ \mu^-$ decay.  The upper limit on $\tau$ in this region comes from having too low $N_{\rm eff}$ already before BBN. But If the decay into muons dominates we will rather have a high $N_{\rm eff}$  
because the amounts of energy released in electrons and in neutrinos by muon decay $\mu\to e \bar \nu\nu$ are similar.
Since data favours values larger than the standard $N_{\rm eff}=3$ the $N_{\rm eff}$ bound will relax somehow. We do not expect them it to disappear because ALPs can still produce too many neutrinos. Also in this case the bound on deuterium should come from a too high D/H, which is less conservative a constraint. 
In any case the bound from He will stay since it mainly comes from a high $\eta_{\rm BBN}$ and the ALP contribution to the expansion at the freeze out of $p\leftrightarrow n$ weak reactions. 

Finally, for $m_\phi>2 m_{\pi^+}$ the most stringent bound comes from $^4$He overproduction due to the presence of charged pions before BBN enhancing the neutron/proton ratio. As we commented this bound does depend very little of the details and branching ratios of the ALP since only a minimal number of pions would do the job. Therefore we expect it not to change very much. However, then quoting this constraint in the $m_\phi-g$ plane this bound would display lower than in the case where only the two photon case is considered.
Only in this region the coupling to two gluons can affect the ALP decay and will certainly increase the pion multiplicity of the decay making the bound on $\tau$ slightly better. The decay into muons can dominate if $m_\phi$ is not too far from $2m_\mu$ and all said in the above paragraph holds. 
It appears that the helium bound will still be the most relevant in this case.

\section{Conclusions}\label{sec:end}

In this paper we have evaluated the impact of pseudo Nambu Goldstone bosons featuring a coupling to two photons, usually called axion-like particles (ALP) and in particular those more massive than the QCD axion. This paper reviews and complements the previous work of Mass\'o and Toldr\`a~\cite{Masso:1995tw,Masso:1997ru}.

ALPs are efficiently created in the early universe via the Primakoff effect and then can decay into two photons leaving traces in the density of neutrinos, the primordial abundances of light nuclei (BBN), the spectrum of the cosmic microwave background or simply creating a diffuse photon background. 
We have found that the more stringent constraints for early decays (before recombination) are set by the density of neutrinos ($N_{\rm eff}$) and the primordial abundances of deuterium and helium, depending on the ALP mass. These bounds are summarised in Fig.~\ref{summa}.
The diffuse photon background and the CMB spectrum are the most relevant for longer lifetimes. 
Interestingly, we have found that these bounds are only slightly modified in scenarios where the ALP has other couplings and other possible decay channels.

At small masses and large couplings, these bounds are complemented by stellar evolution arguments, particularly the ratio of red giants to horizontal  branch stars in globular clusters, which we have revisited to define precisely the high mass frontier. 
Altogether, these arguments exclude a huge patch of parameter space, shown in our summary plots Figs.~\ref{fig:bounds} and~\ref{fig:boundsLife}.

\section*{Acknowledgements}
We are indebted to Georg Raffelt for proposing the \emph{putatively} small project that turned into this work. 
We also want to thank Sara Cavallin and J\"org J\"ackel for good advice regarding the aesthetics of the paper and Daniel Greenwald for reading the manuscript.
\appendix

\section{The globular cluster bound for high mass ALPs }\label{sec:astro}

The ratio of red giants (RG) to horizontal branch (HB) stars in globular clusters is an indicator of the presence of weakly coupled low mass particles which can efficiently drain energy from the stellar cores~\cite{Raffelt:1996wa}. The non standard energy loss prolongs the RG phase and shortens the HB in a different way, being these effects sensitive to the type of particle and coupling to matter. For particles coupled to two photons, the lifetime of HB stars is more affected than the RG. Qualitative studies~\cite{Raffelt:1987yb,Raffelt:1996wa} show that the exotic energy emitted per unit time and mass, averaged over a typical HB core, has to satisfy
\be
\label{HBcons}
\langle\epsilon_{\rm exotic}\rangle \lesssim 10\, {\rm g}^{-1}\, {\rm erg}\, {\rm s}^{-1}
\ee

The main reaction producing ALPs in a HB core through the two photon coupling is the Primakoff conversion in the Coulomb field of ions in the plasma, i.e. $\gamma q\to \phi q$ where the $q$s are mostly He$^{2+}$ and protons. 
The cross section in the ion rest frame can be written as  
\bea
\sigma_{\gamma q}(\omega) &=& \frac{\alpha g^2 Q^2}{8} \[\(1+\frac{k_s^2}{4 \omega^2}-\frac{m_\phi^2}{2 \omega^2}\)\log\(\frac{2\omega^2(1+\beta)+k_s^2-m_\phi^2}{2\omega^2(1-\beta)+k_s^2-m_\phi^2}\)-\beta\right. \\
&&\left. 
\quad\quad -\frac{m_\phi^4}{4k_s^2 \omega^2}
\log \(\frac{m_\phi^4+k_s^2\(2\omega^2(1+\beta)-m_\phi^2 \)}
{m_\phi^4+k_s^2\(2\omega^2(1-\beta)-m_\phi^2 \)}\) \] 
\eea
where $\beta=\sqrt{\omega^2-m_\phi^2}/\omega$ is the ALP velocity, we have assumed the ALP mass much smaller than the mass of the ion, so that the ALP energy is equal to the initial photon energy $\omega$, 
and we have taken into account Coulomb screening following~\cite{Raffelt:1985nk}. In an non-degenerate plasma, the screening scale is given by the Debye-Hueckel formula
\be
k_s^2= \frac{4\pi \alpha}{T}n_q
\ee 
where the sum involves all charged species in the plasma. 
At low masses, the logarithmic divergence of the cross section (inherited from the well known forward divergence of Coulomb scattering) is cut-off by screening. However, when considering large masses, it is the ALP mass which plays this role and screening becomes irrelevant.

\begin{figure}[tbp] 
   \centering
      \includegraphics[width=2.9in]{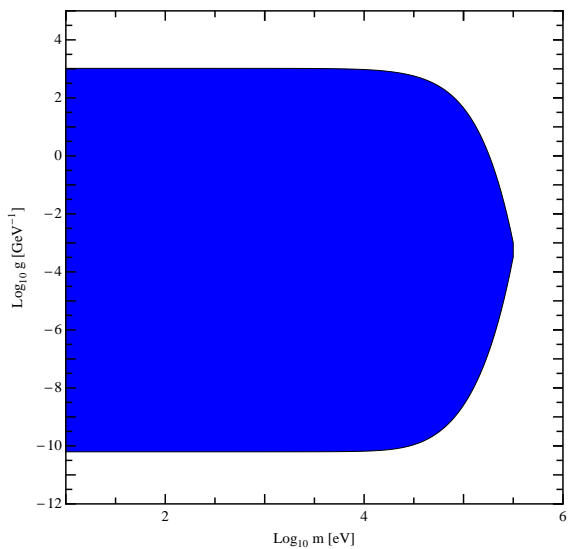} 
   \caption{Bounds on ALPs parameter space from energy loss in Horizontal Branch stars.}
   \label{fig:pike}
\end{figure}

The energy loss per unit mass is given by folding the cross section with the photon phase space and thermal phase space distribution $f_{\rm BE}(\omega)=(e^{\omega/T}-1)^{-1}$
\be
\epsilon = \frac{1}{\rho}\int \frac{d^3{\bf k}}{(2\pi)^3}\omega f_{\rm BE}(\omega) \sum Q_j^2 n_j \sigma_{\gamma q}(\omega)\, . 
\ee
For typical conditions in a HB core, $T \sim 8.6$ keV and matter density $\rho=10^4$ g cm$^{-3}$, the screening scale is 
$k_s\sim 27$ keV. 
 
The energy loss grows with $g$ and so does the impact on the star. However, at some point $g$ can be so large that the
inverse Primakoff process, $\phi q\to \gamma q$, cannot be neglected and ALPs will be reabsorbed inside the star. 
This does not invalidate the constrain, since ALPs will nevertheless transfer energy from the core to the external shells of the star which has a  very similar effect on the star evolution~\cite{Raffelt:1988rx}.  However, if the ALP mean free path, $\lambda$, gets very small, even the energy transfer will become eventually negligible. Unfortunately, the energy transfer on a HB core is convective and therefore it is difficult to ascertain for which exact value of $g$ ALPs are harmless. In order to get an estimate, the authors of~\cite{Raffelt:1988rx} propose to compare the ALP energy transfer with the radiative energy transfer mediated by photons.  Since radiative transfer is subleading in HB cores, the limit of validity of the bound is conservative. Thus, we impose that the contribution of ALPs to the Rooseland mean opacity is smaller than the standard value in a HB core $\kappa_0\sim 0.5$ cm$^2/$g, which means 
\be
\kappa_\phi\equiv \frac{\int_m^\infty \beta^2 f_{\rm BE}'(\omega)\omega^3d\omega}{\rho \int_m^\infty \lambda_\omega \beta^2 f_{\rm BE}'(\omega)\omega^3d\omega} > \kappa_0\,  , 
\ee 
where $f_{\rm BE}'=\partial f_{\rm BE}/\partial T$, the mean free path is given by $\lambda_\omega^{-1}=\sum_j Z_j^2 n_j \sigma_{\phi Z}(\omega)$ with  the inverse Primakoff cross section given by detail balance $\sigma_{\phi Z}(\omega)=\sigma_{\gamma Z}(\omega)/(2\beta^2)$. 
The full range excluded by HB stars in globular clusters is shown in Fig.~\ref{fig:pike} and is also reproduced Figs.~\ref{fig:bounds} and~\ref{fig:boundsLife}.


\end{document}